\documentclass[fleqn,usenatbib]{mnras}

\usepackage{newtxtext,newtxmath}

\usepackage[T1]{fontenc}

\DeclareRobustCommand{\VAN}[3]{#2}
\let\VANthebibliography\thebibliography
\def\thebibliography{\DeclareRobustCommand{\VAN}[3]{##3}\VANthebibliography}

\usepackage{graphicx}	
\usepackage{amsmath}
\usepackage{amssymb}	
\usepackage{longtable}
\usepackage{threeparttable}
\usepackage{dcolumn}
\usepackage{multirow}
\usepackage{booktabs}
\usepackage{array}

\title[SDSS J0812]{New evidence for the precession of tilted disk in SDSS J081256.85+191157.8}

\author[Sun et al.]{
		Qi-Bin Sun$^{1,2,3,4}$, 
	Sheng-Bang Qian$^{1,2,3,4}$\thanks{E-mail: qsb@ynao.ac.cn},
	Li-Ying Zhu$^{1,3,4}$,
	Wen-Ping Liao$^{1,3,4}$,
	Er-Gang Zhao$^{1,3}$,
	\newauthor{
		Fu-Xing Li$^{1,3}$,
		Xiang-Dong Shi$^{1,3}$},
	    Min-Yu Li$^{1,3,4}$
	\\
	$^{1}$Yunnan Observatories, Chinese Academy of Sciences, Kunming 650216, China\\
	$^{2}$ Department of Astronomy, Key Laboratory of Astroparticle Physics of Yunnan Province, Yunnan University, Kunming 650091, China\\
	$^{3}$Center for Astronomical Mega-Science, Chinese Academy of Sciences, 20A Datun Road, Chaoyang District, Beijing, 100012, China\\
	$^{4}$University of Chinese Academy of Sciences, No.1 Yanqihu East Rd, Huairou District, Beijing 101408, China}
\date{Accepted 2023 June 18. Received 2023 June 11; in original form 2023 March 21}
\pubyear{2023}

\begin{document}
\label{firstpage}
\pagerange{\pageref{firstpage}--\pageref{lastpage}}
\maketitle


\begin{abstract}
	Super-orbital signals and negative superhumps are thought to be related to the reverse precession of the nodal line in a tilted disk, but the evidence is lacking. Our results provide new evidence for the precession of the tilted disk. Based on the \textit{TESS} and \textit{K2} photometry, we investigate the super-orbital signals, negative superhumps, positive superhumps, and eclipse characteristics of the long-period eclipsing cataclysmic variable star SDSS J0812. We find super-orbital signals, negative superhumps, and positive superhumps with periods of 3.0451(5) d, 0.152047(2) d, and 0.174686(7) d, respectively, in the \textit{K2} photometry, but all disappear in the \textit{TESS} photometry, where the positive superhumps are present only in the first half of the same campaign, confirming that none of them is permanently present in SDSS J0812. In addition, we find for the first time a cyclic variation of the O-C of minima, eclipse depth, and negative superhumps amplitudes for 3.045(8) d, 3.040(6) d, and 3.053(8) d in SDSS J0812, respectively, and all reach the maximum at $\sim$ 0.75 precession phases of the tilted disk, which provides new evidence for the precession of the tilted disk. We suggest that the O-C and eclipse depth variations may come from a shift of the brightness center of the precession tilted disk. 
	Our first finding on the periodic variation of negative superhumps amplitude with the super-orbital signals is significant evidence that the origin of negative superhumps is related to the precession of the tilted disk.

\end{abstract}




\section{Introduction} \label{sec:intro}

Cataclysmic variable stars (CVs) are semi-detached close binary systems consisting of a white dwarf (primary) and a late-type main sequence star (secondary). The secondary star fills its Roche lobe and transfers mass via the L1 inner lagrange point to the region around the primary star \citep{warner1995cat}. CVs have two superhump modulations: negative superhumps (NSH) and positive superhumps (PSH) with periods a few percent smaller and larger than the orbital period.
PSH are thought to be caused by the line of apsides precession due to the 3:1 tidal resonance between the accretion disk and the secondary star during the superoutburst (e.g., \citealp{vogt1982z}; \citealp{osaki1985irradiation}; \citealp{wood2011v344}). PSH are commonly found in the CVs subtype SU UMa.
The theory of the NSH is not as complete as that of the PSH, and current research suggests that the NSH arises from the action between the reverse precession of the nodal line of the tilted disk and the orbital motion of the binary star (e.g., \citealp{1985A&A...143..313B}; \citealp{patterson1999permanent}; \citealp{harvey1995superhumps}; \citealp{sun2022study}). Nevertheless, there is no consensus on why the disk is tilted, why it reverses, and how it acts with the orbital motion. However, NSH can be reproduced based on a tilted disk (e.g., \citealp{wood2009sph}; \citealp{Montgomery2009MNRAS}; \citealp{2021PASJ...73.1225K}). NSH are found not only in CVs but also in low-mass X-ray binaries \citep{retter2002detection}. According to \cite{Barrett1988MNRAS}, the NSH may result from an accretion stream impacting and sweeping across the face of a tilted disk. \cite{Patterson1997PASP} suggests that the source of the NSH is gravitational energy and that if the disk is tilted, the stream can easily crash into the inner disk to release more energy. \cite{Wood2007ApJ...661.1042W} proposes that the NSH is caused by hot spots periodically sweeping across the surface of the tilted disk and that the energy released by the stream impulse is also periodically changing due to the periodic change in the location of the hot spots.

In NSH systems, in addition to the NSH, there is generally a super-orbital signal (SOR) of about several days, whose period can be described as:
\begin{equation} \label{eq:pso}
	\dfrac{1}{P_{\rm prec}}  = \dfrac{1}{P_{\rm sor}} =\dfrac{1}{P_{\rm nsh}} - \dfrac{1}{P_{\rm orb}}  
\end{equation} 
The SOR is generally considered to be caused by the reverse precession of the line of nodes of the tilted accretion disk (e.g., \citealp{Katz1973NPhS..246...87K}; \citealp{Barrett1988MNRAS}; \citealp{harvey1995superhumps}). The three-dimensional smoothed particle hydrodynamic(3D-SPH) simulations of \cite{Montgomery2012} suggest that the disk can tilt naturally without the influence of magnetic fields or other radiation sources. This tilt is attributed to the lift generated by the accretion stream colliding with the edge of the disk, which causes the stream to be spilled in a different direction at the top and bottom of the tilted disk. When the accretion disk is slightly tilted, the impact point of the accretion stream changes slightly up and down, resulting in more stream being spilled into the inner ring of the disk after the impact on the edge of the disk.
\cite{2013MNRAS.435..707A} found that the eclipse depth of the CV AQ Men is highly variable and suggested that the eclipse depth could be used to study the systematic variations of the tilted disk. 
Inspired by this, \cite{2021MNRAS.503.4050I} quantified the eclipse depth of AQ Men based on \textit{TESS} photometry. They found that the eclipse depth of AQ Men showed a cyclical variation consistent with the SOR cycle, providing evidence for the tilted disk theory.
Similarly, \cite{2017Boyd} also found a cycle of change in the eclipse depth, width and skew with SOR for the nova-like variable DW UMa.
However, the evidence that the NSH and SOR originate from the reverse precession of the nodal line in a tilted disk is still lacking, and further observations and tests are needed. This paper's primary purpose and contribution are to portray the NSH and SOR from different perspectives and provide new evidence.

SDSS J081256.85+191157.8 (hereafter SDSS J0812) is a long-period ($P_{\rm orb}$ $\sim$ 3.84 hr) eclipsing nova-like object, first observed by the Sloan Digital Sky and identified as a nova-like object by the obtained spectra \citep{Szkody2006}. \cite{2014NewA...28...49G} analyzed SDSS J0812 photometry and obtained an orbital period of 0.160151(79) d and found an NSH of 0.148159(86) d.  
The photometric and spectroscopic analysis performed by \cite{2015AJ....149..128T} on SDSS J0812 and corrected the orbital period to 0.1600(2) d, again finding the NSH close to 0.148159 d. This paper analyzes the NSH, PSH, and SOR of SDSS J0812 based on the \textit{TESS} and \textit{K2} photometry, providing new evidence for the precession of the tilted disk.

This paper is structured as follows. Section \ref{sec:style} introduces the \textit{K2} and \textit{TESS} photometry of SDSS J0812. 
In Section \ref{sec:Analysis}, the data of K2(c18), K2(c05), and \textit{TESS} are analyzed separately, mainly including the analysis of SOR, NSH, PSH, O-C and the change of eclipse depth.
In section \ref{sec:DISCUSSION}, we calculate some of the parameters of SDSS J0812 and discuss the significance of the discovered new phenomenon. Section \ref{sec:CONCLUSIONS} is the summary.


\section{\textit{K2} and \textit{TESS} Photometry } \label{sec:style}

We used the data from the \textit{Kepler Space Telescope K2 Missions} \citep{borucki2010kepler,howell2014k2} and the space telescope \textit{Transiting Exoplanet Survey Satellite} (\textit{TESS}, \citealp{Ricker2015journal}).
K2 has two modes of observation: long cadence (30 minutes) and short cadence (1 minute). During Campaigns 5 (hereafter K2(c05)) and 18 (hereafter K2(c18)), \textit{K2} observed SDSS J0812 in short cadence (1 minute) mode, with an exposure time of 60 s and a 420 to 910 nm wavelength range. See Table \ref{tab:LOG} for the observation period. 

\begin{figure}
	\includegraphics[width=\columnwidth]{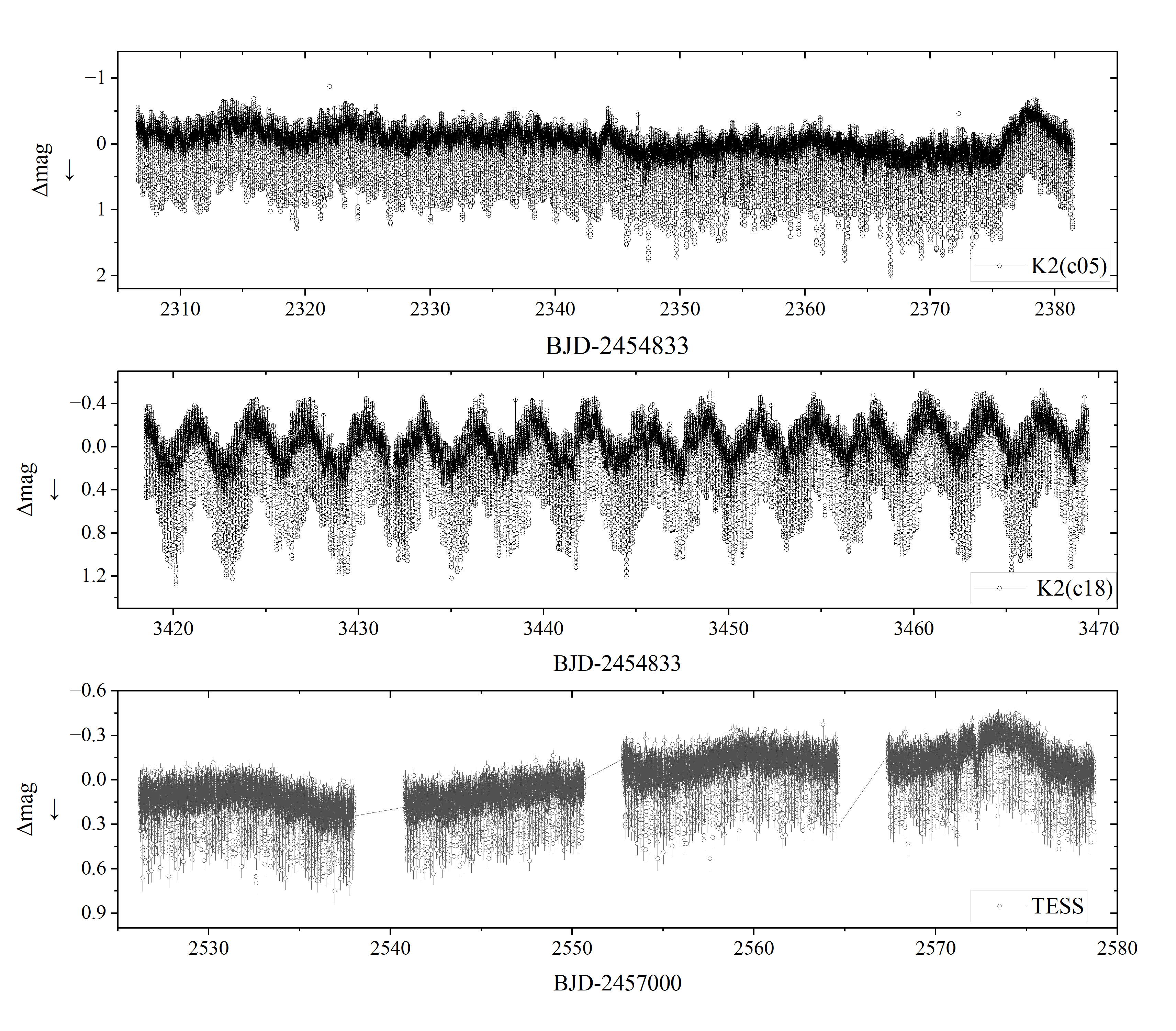}
	\caption{The light curves of K2(c05), K2(c18) and \textit{TESS}. \label{fig:alllight}}
\end{figure}

\textit{TESS} in sectors 44, 45, and 46 observed SDSS J0812,  with exposure times of 120 s at wavelengths from 600 to 1000 nm. The span of the observations is shown in Table \ref{tab:LOG}.
Data were downloaded from the Mikulski Archive for Space Telescopes (MAST)\footnote{https://mast.stsci.edu/} and
the ExoFOP-\textit{TESS} webpage\footnote{https://exofop.ipac.caltech.edu/tess/target.php?id=172518755}.
The light curves for \textit{K2} and \textit{TESS} are passed on to the Pre-Search Data Conditioning (PDC, \citealp{2010SPIE.7740E..1UT}) to correct for systematic errors. In our work, PDC fluxes are used.
We use the same treatment as for \cite{2013PASJ...65...95O} and \cite{2020PASJ...72...94K}, converting the flux to magnitude using Mag = -2.5lg(Flux) and used the mean magnitude as the 0 points (see Fig. \ref{fig:alllight}).

\begin{table}
	\centering
	\caption{Journal of observations. \label{tab:LOG}}
	\setlength{\tabcolsep}{1mm}{
		\begin{tabular}{ccccc} 
			\toprule
			Telescope & Sectors/Campaigns& Start&Start&End\\
			 &  &  UT  &MBJD & MBJD\\
			\midrule
	\textit{K2}&c05&2015-04-27&57139.09597&57213.94425\\
&c18&2018-05-13&58251.03105&58301.91072\\
\textit{TESS} &s44&2021-10-12&59499.70230&59523.94081\\
&s45&2021-11-07&59525.22565&59550.12787\\
&s46&2021-12-03&59551.49739&59578.20590\\
			\bottomrule
	\end{tabular}}
\end{table}

\begin{table*}
	\centering
	\caption{Results of the analysis of \texttt{Period04}.\label{tab:freq}}
	\setlength{\tabcolsep}{2mm}{
		\begin{tabular}{lllllll} 
			\toprule
			Sectors/Campaigns  &  Name &  frequency &  period&  amplitude& noise&  SNR\\
		 &  & [1/d] & [d]& [mag or d]& [mag or d] & [mag or d]\\
		 	\midrule
	K2(c05)&f2&4.0773(3)&0.24526(2)&0.027(1)&0.00364 &7.52\\
&f3&5.7245(2)&0.174686(7)&0.035(1)&0.00476 &7.41\\
&f4&6.24797(9)&0.160052(2)&0.088(1)&0.00543 &16.29\\
K2(c18)&f1&0.32840(5)&3.0451(5)&0.1775(8)&0.01640 &10.82\\
&f2&4.0771(8)&0.24527(5)&0.016(1)&0.00270 &6.06\\
&f4&6.2479(1)&0.160053(3)&0.0964(8)&0.00899 &10.72\\
&f5&6.57693(9)&0.152047(2)&0.0891(8)&0.00907 &9.82\\
\textit{TESS}&f4&6.2478(3)&0.160057(7)&0.050(1)&0.00402 &12.54\\
K2(c18)&$\mid A \mid$&0.3276(9)&3.053(8)&0.029(3)&0.00393 &7.28\\
&Eclipse depth&0.3289(7)&3.040(6)&0.096(6)&0.01138 &8.43\\
&O-C&0.3284(9)&3.045(8)&0.00026(2)&0.00003 &7.53\\
		\bottomrule
\end{tabular}}
\end{table*}

\section{Data Analysis and Results} \label{sec:Analysis}

\subsection{Signals} \label{subsec:Frequencies}

\begin{figure*}
	\includegraphics[width=1.5\columnwidth]{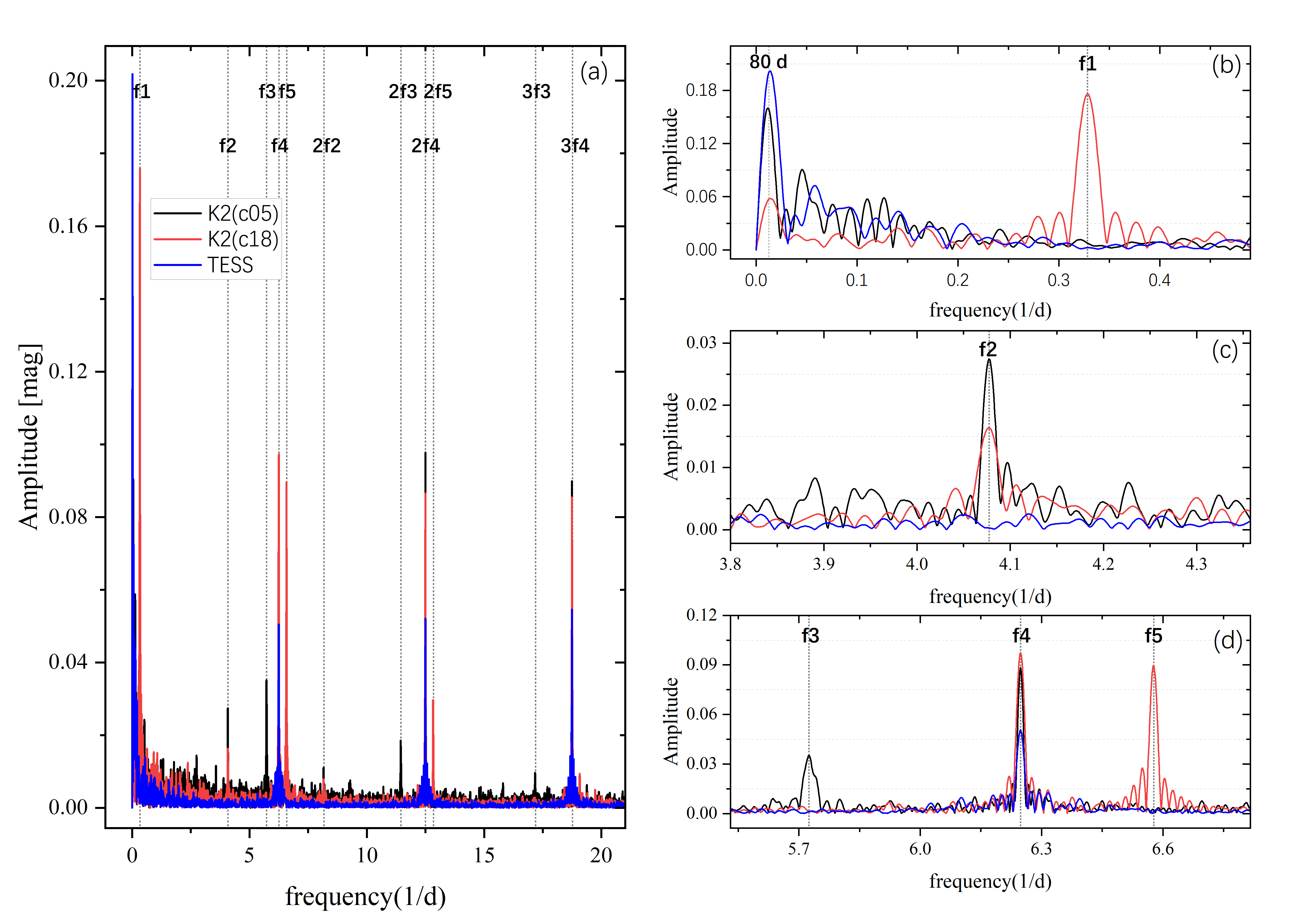}
	\caption{Frequency-Amplitude spectrum of SDSS J0812. The black, red, and blue solid lines correspond to the K2(c05) analysis results, K2(c18), and \textit{TESS}. (b), (c), and (d) are enlarged plots of the different frequencies.\label{fig:allpower}}
\end{figure*}

By dividing the data into \textit{TESS}, K2(c05), and K2(c18) for time-varying analysis and using \texttt{Period04} (see \citealp{Period04} for details) to calculate the frequency, amplitude, uncertainty, noise, and signal-to-noise ratio (SNR) of the three data components. We plotted the frequency-amplitude spectrum and labeled the main frequencies and their harmonics in Fig. \ref{fig:allpower}. The results revealed the presence of five frequency components in SDSS J0812 (see Table \ref{tab:freq}). In the K2(c05) data, signals with frequencies of 4.0773(3) d$^{-1}$, 5.7245(2)~d$^{-1}$ and 6.24797(9) d$^{-1}$ were present. Signals with frequencies of 0.32840(5)~d$^{-1}$, 6.2479(1) d$^{-1}$ and 6.57693(9) d$^{-1}$ were observed in K2(c18). Only a significant signal of 6.2478(3) d$^{-1}$ was detected in the \textit{TESS} data. However, signals similar to the data spanning about 80~d were not counted. In the next work, the data of each segment will be analyzed separately to verify the possible existence of periodic signals.

\subsection{K2(c18)} \label{subsec:K2(c18)1}
\subsubsection{Super-Orbital Signals and Negative Superhumps} \label{subsec:SO and NSH}

Because the periodic signal of K2(c18) is visible to the naked eye, the study will start with the light curve of K2(c18). 
In K2(c18) there are periodic signals with periods 0.160053(3) d, 0.152047(2) d and 3.0451(5) d respectively. Based on \cite{2014NewA...28...49G} and the eclipse, it can be determined that 0.160053(3) d is the orbital period, 0.152047(2) d is the NSH, 3.0451(5) d is the SOR. 
The next step will be individually verifying SOR and NSH in the light curve. The least-squares linear superimposed sine model without weights is first used to fit the out-of-eclipse curve (the solid blue circle for curve \#1 in Fig. \ref{fig:all_light_k22} d):
\begin{equation} \label{eq:sine}
	Magnitude(BJD) = Z + B*BJD + A*\sin(2\pi*(freq*BJD+p))
\end{equation}
$Z$, $B$, $A$, $freq$, and $p$ are the fitted intercept, slope, amplitude, frequency, and phase, respectively.
The best-fit results are shown in Table \ref{tab:sine}, with errors in each fitted parameter derived from the covariance. The best-fit frequency was 0.32851(4) d$^{-1}$ (3.0440(4) d) in agreement with the 3.0451(5) d.

\begin{figure}
	\includegraphics[width=\columnwidth]{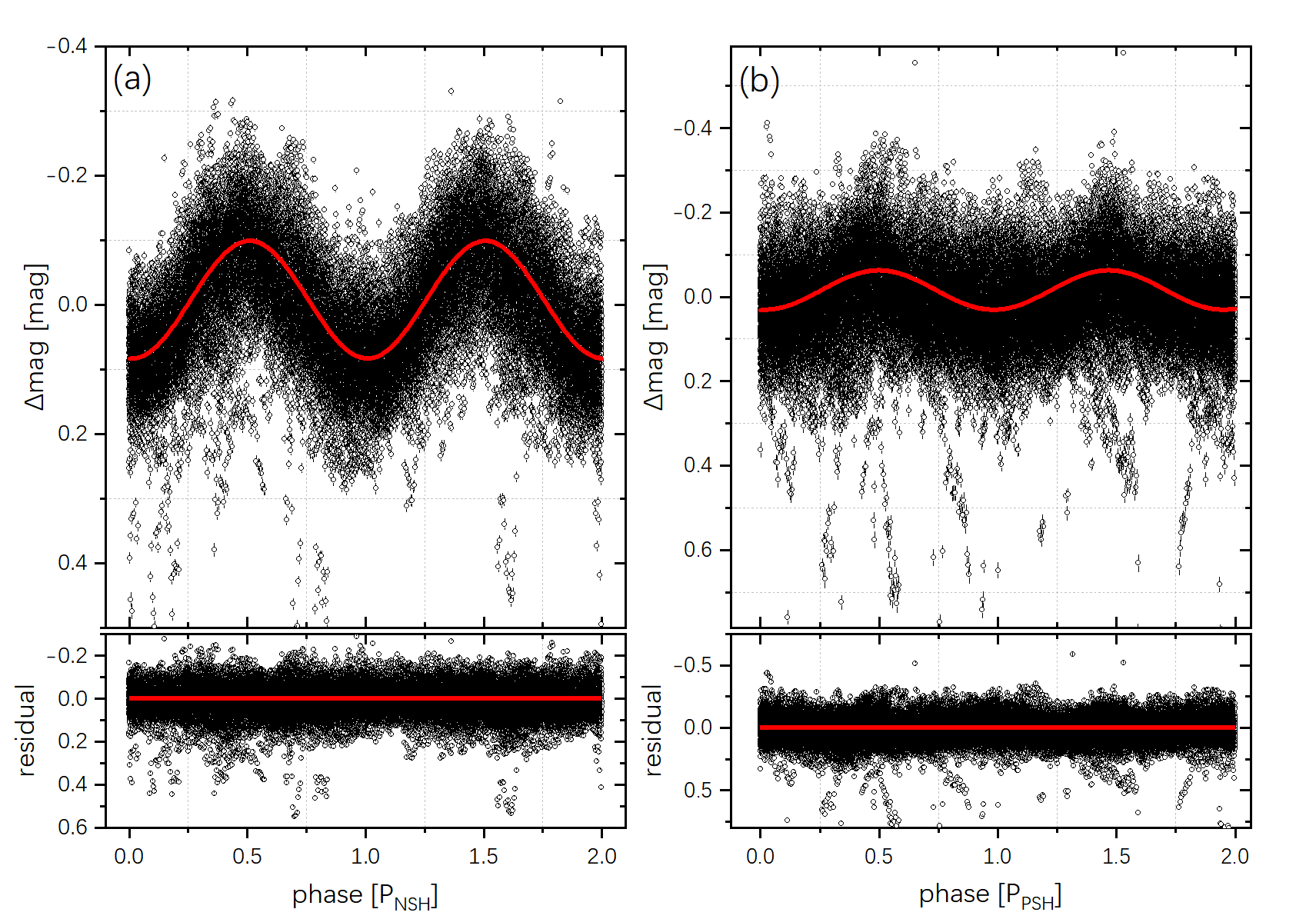}
	\caption{Folded diagram of NSH and PSH. (a): Phase folding diagram of NSH with a folding period of 0.152047(2) d; (b): Phase folding diagram of PSH with a folding period of 0.174686(7) d. The red curves are the sine fit. \label{fig:nsh}}
\end{figure}

We get the NSH of SDSS J0812 being longer than 0.148159(86) d obtained by \cite{2014NewA...28...49G}, and the NSH excess ($\epsilon ^- = (P_{\rm orb} - P_{\rm nsh})/P_{\rm orb}$ ) was calculated to be $\epsilon ^- \sim$ 0.050.
To obtain more information on NSH, we performed a segmented fit using Eq. \ref{eq:sine}. The data used are the residuals of the SOR theoretical curve to all data (the solid blue curve \#2 in Fig. \ref{fig:all_light_k22} d).
We split the data from the detrended SOR out-of-eclipse curve of K2(18) into 80 segments of approximately 0.64 d each (some segments lasted slightly longer or shorter than 0.64 d due to gaps in some of the data, see Table \ref{tab:k2(c18)}).
The best-fit results are shown in Table \ref{tab:k2(c18)} and Fig. \ref{fig:all_light_k22} d. The average frequency of the fit was 6.51(3) d$^{-1}$. 
Based on the results obtained by fitting to the out-of-eclipse curve, the linear part of the NSH was removed, and the NSH profile became more visible by folding with 0.152047(2) d (see Fig. \ref{fig:nsh} a).

The most valuable information obtained in the segment fitting process is the amplitude of NSH.
The absolute value of the amplitude ($\mid A\mid$) was used as the y-value, and the average BJD of the individual segments was plotted for time (see Fig. \ref{fig:all_light_k22} c). 
The results show that there is a periodic variation in the amplitude of the NSH, whose period was obtained as 3.053(8) d using \texttt{Period04} (see Fig. \ref{fig:f-4} d), and also using Eq. \ref{eq:sine} to fit, the results of the fit are shown in Table \ref{tab:sine}. To show the variation of the NSH amplitude more visually, we remove the linear part obtained by segmental fitting of the out-of-eclipse curve \#2 in Fig. \ref{fig:all_light_k22} d (see Fig. \ref{fig:1}). 

Our study shows that a sinusoidal fit can verify SOR and NSH signals to the curve. The very peculiar result is a periodic variation in the amplitude of the NSH in SDSS J0812 with a period close to the SOR.

\subsubsection{Eclipse depth and O-C} \label{subsec:O-Ck22}

\begin{figure*}
	\includegraphics[width=1.5\columnwidth]{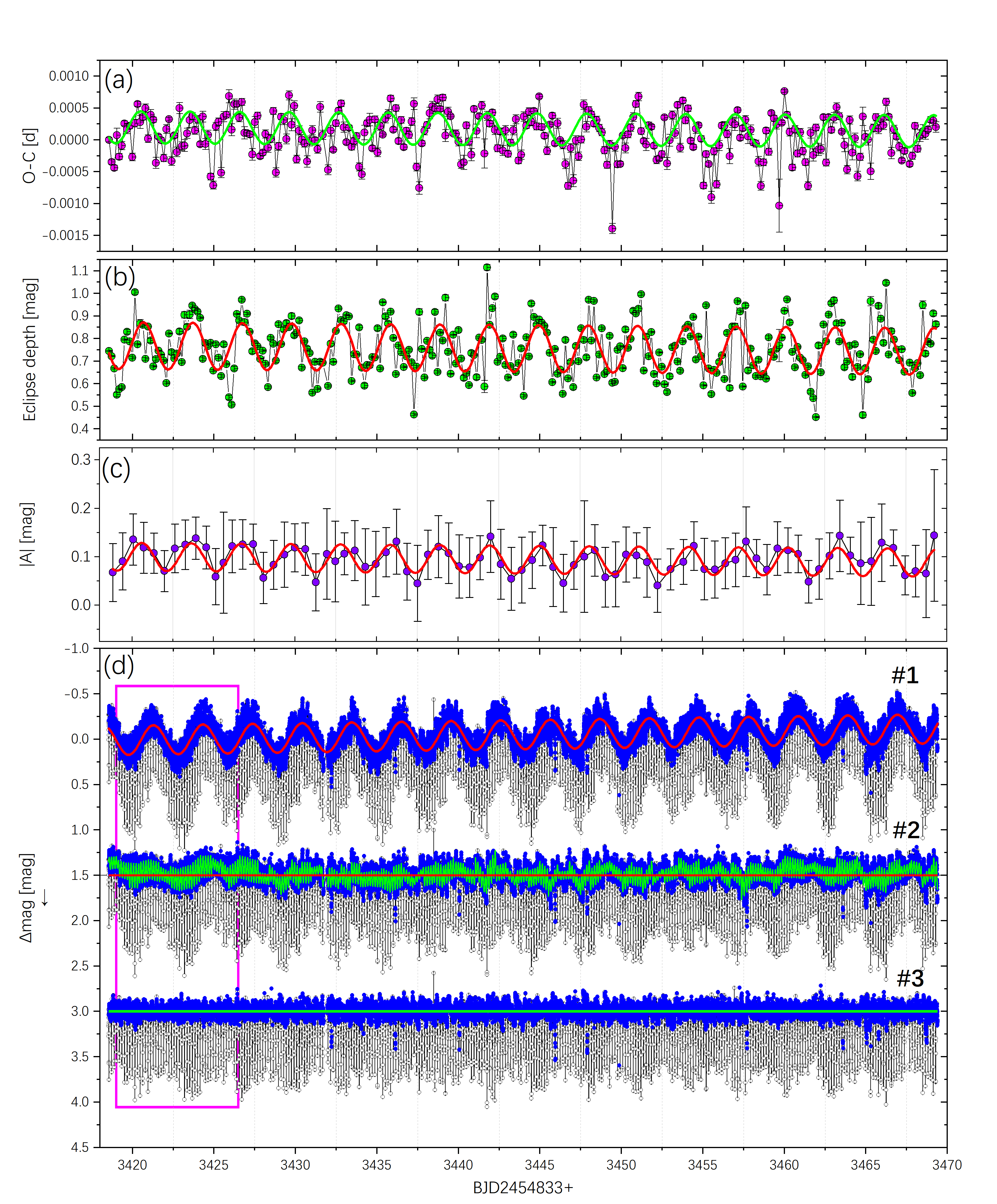}
	\caption{A combined plot of the analytical process and results for K2(c18). (a): O-C plots; (b): plots of the eclipse depth versus time; (c): plots of the absolute value of the NSH amplitude; the solid lines in plots (a), (b), and (c) are the results of the fit to Eq. \ref{eq:sine}. (d): we split into 3 main curves; \#1: all light curves of K2(c18), the solid blue circle is the out-of-eclipse part, the red curve is a linear superimposed sine fit to the out-of-eclipse curve; \#2: the residuals of the \#1 fit to all curves plus 1.5 mag, the green curve is a segmented linear superimposed sine fit to the out-of-eclipse curve; \#3: the residuals of the segmented linear superimposed sine fit to all curves plus 3 mag. \label{fig:all_light_k22}}
\end{figure*}

\begin{figure*}
	\centering
	\includegraphics[width=1.5\columnwidth]{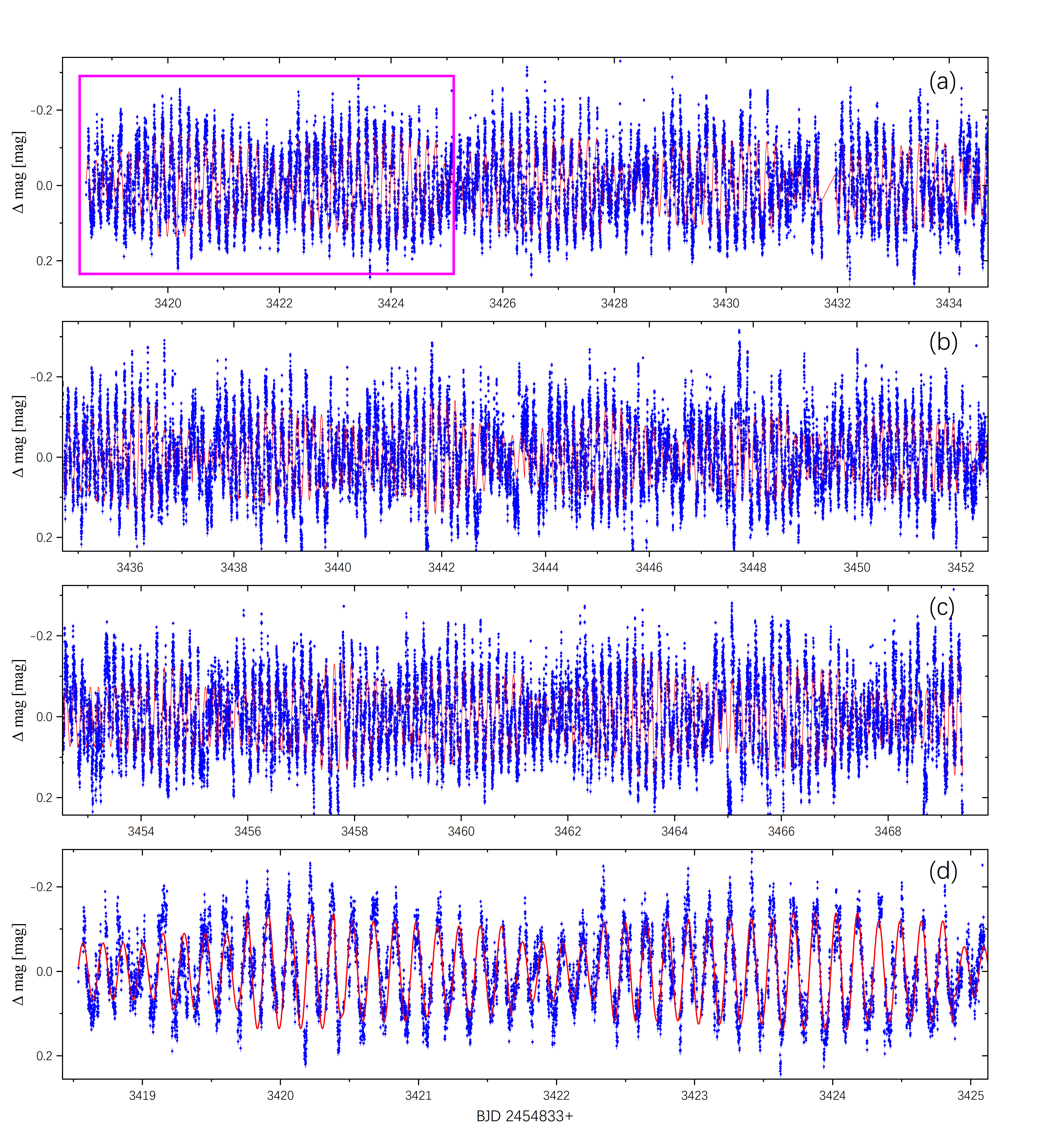}
	\caption{Plots of NSH amplitude variation. (a), (b) and (c) panels are the out-of-eclipse curve \#2 with the linear trend removed in Fig. \ref{fig:all_light_k22} d, and the red solid line is a linear superimposed sinusoidal fit with the linear part of the segment removed; (d) is a zoomed-in view of the magenta rectangular box in panel (a). \label{fig:1}}
\end{figure*}

The next step will be to analyze the eclipse depth and the minima for SDSS J0812. We will need to use the out-of-eclipse curve and the minima. To exclude the influence of SOR and NSH, the light curve with the SOR and NSH trends removed is used (the curve \#3 in Fig. \ref{fig:all_light_k22} d). 

In Section \ref{subsec:SO and NSH}, we have already performed two separate linear superimposed sine fits the out-of-eclipse curves. The first one was fitted to SOR to obtain the theoretical SOR curve (see the red curve of \#1 in Fig. \ref{fig:all_light_k22} d), and all curves were subtracted from the theoretical SOR curve so that the light curve with the SOR trend removed was obtained (see the curve \#2 in Fig. \ref{fig:all_light_k22} d).
For the second fit, we split the out-of-eclipse curves with the SOR trend removed into 80 segments and performed a linear superimposed sine fit to each to obtain the theoretical NSH curve (see the green curve of \#2 in Fig. \ref{fig:all_light_k22} d).
We used the theoretical NSH to take the residuals for all curves, so we ended up with light curves with SOR and NSH removed (see the curve of \#3 in Fig. \ref{fig:all_light_k22} d).

\begin{figure}
	\includegraphics[width=\columnwidth]{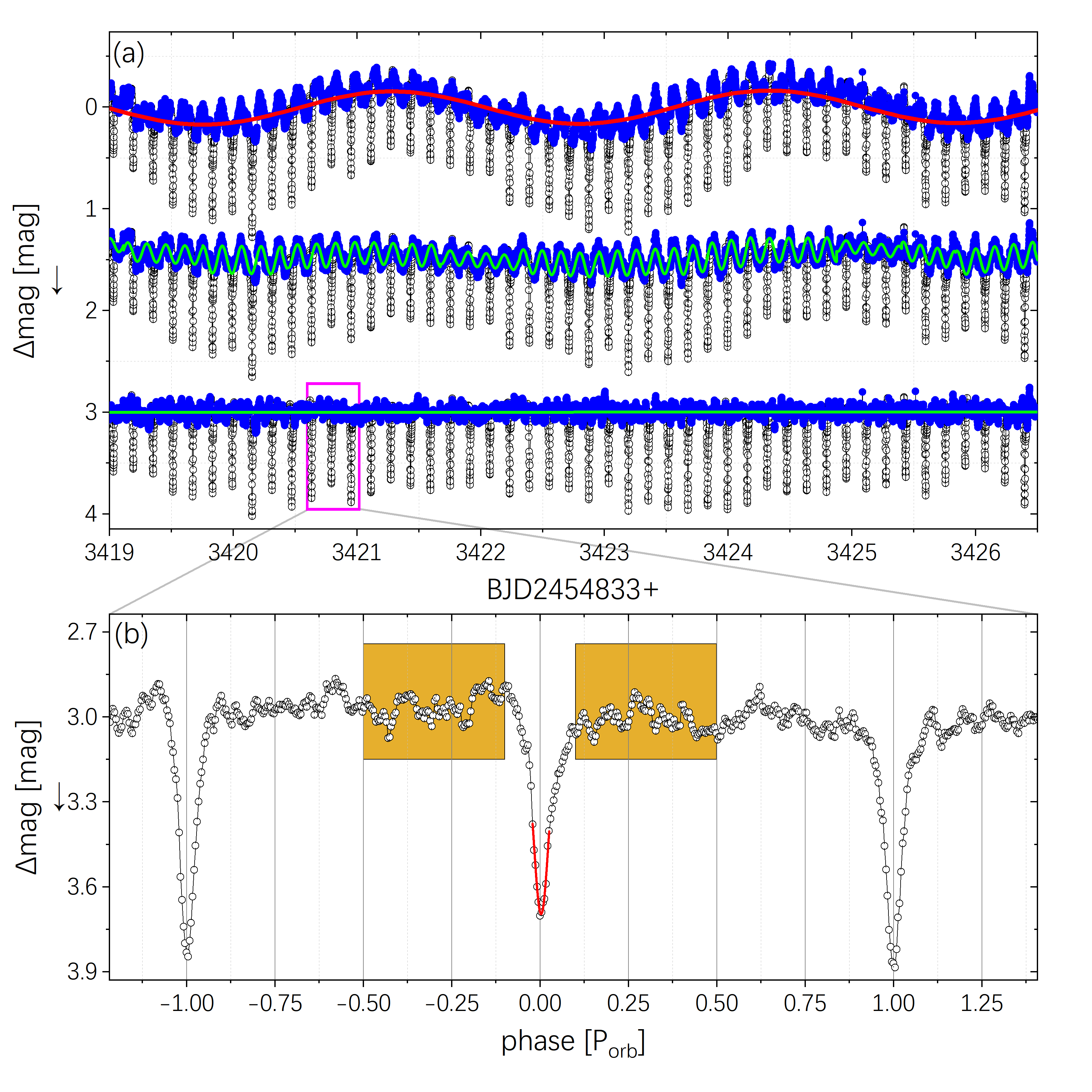}
	\caption{Examples of linear superimposed sinusoidal fitting, Gaussian fitting, and calculation of eclipse depth. (a): a zoomed-in view of the magenta rectangular box in Fig. \ref{fig:all_light_k22} d, where the solid green line is a linear superimposed sinusoidal fit of the segments, and the obvious segmentation trace can be seen; (b): a zoomed-in view of the magenta rectangular box of the plate (a) but with the transverse axes converted to phase, where the solid red line is a Gaussian fit to a phase width of 0.05, and the brown boxed out-of-eclipses part (-0.5 to -0.1 and 0.1 to 0.5 phases). \label{fig:example_minima}}
\end{figure}

Based on the light curves with SOR and NSH, we were removed, and the Gaussian fit was used to calculate the minima coordinates of the eclipse (see Fig. \ref{fig:example_minima} b). The minima were obtained by fitting the 0.05 phase width at the mid-eclipse time, and the error in time was derived from the covariance. After obtaining the eclipse minima coordinates, we took -0.5 to -0.1 and 0.1 to 0.5 phases of each eclipse as the corresponding out-of-eclipse curve. The difference in magnitude between the mean value of the out-of-eclipse and the minima was taken as the eclipse depth (see Fig. \ref{fig:example_minima} b). 
Table S2 of the Supplementary materials shows the obtained minima and eclipse depths.

\begin{figure}
	\includegraphics[width=\columnwidth]{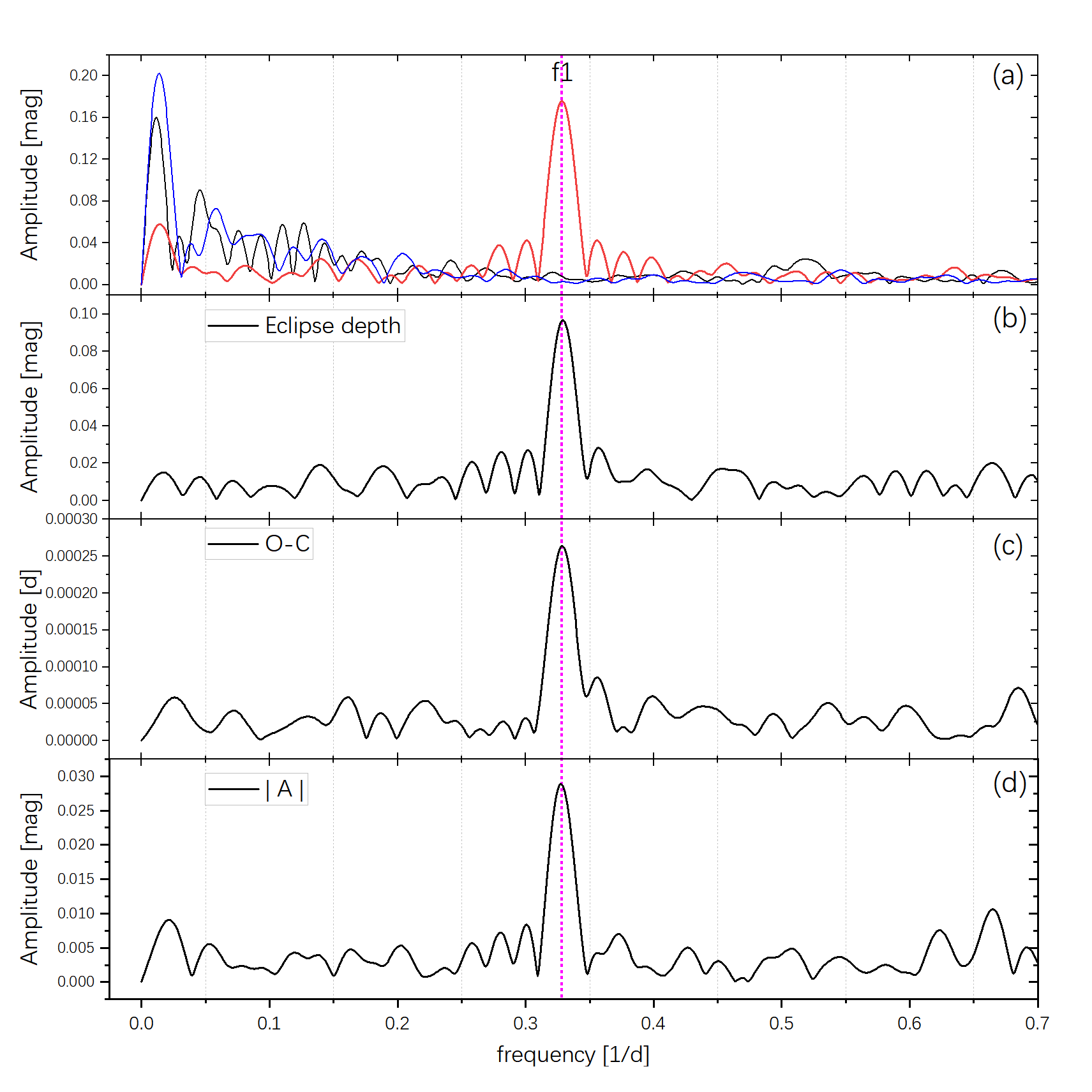}
	\caption{Frequency-Amplitude spectrum of different curves. (a): same as Fig. \ref{fig:allpower} b; (b): frequency-amplitude spectrum of the eclipse depth corresponds to Fig. \ref{fig:all_light_k22} b; (c): spectrum of the O-C corresponds to Fig. \ref{fig:all_light_k22} a; (d): spectrum of the absolute value of NSH amplitude corresponds to Fig. \ref{fig:all_light_k22} c. The vertical magenta dotted line in the figure is 3.0451(5) d. \label{fig:f-4}}
\end{figure}

The 316 minima were obtained by fitting the K2(c18) curve using a Gaussian fit. Using the first minima of K2(c18) as the initial epoch, a new ephemeris with a period of 0.160053(3) d was obtained:
\begin{equation} \label{eq:ephemeris}
	Min.I=BJD2458251.55295(3) + 0.160053(3) \times E 
\end{equation}
E is the number of cycles. For uniformity, all the O-C analyses that follow use this ephemeris.
Based on the new ephemeris, we carried out an O-C analysis of all the minima of K2(c18). The results show the same periodic variation in O-C (see Fig. \ref{fig:all_light_k22} a). The period and amplitude of O-C variation obtained from the frequency-amplitude spectrum are 3.045(8) d and 0.00026(2) d, respectively, which are in general agreement with the SOR period (see Fig. \ref{fig:f-4} c).
By plotting the eclipse depth and time, it can be seen that there is a periodic variation. The period and amplitude of variation are 3.040(6) d and 0.096(6) mag, respectively, obtained by using \texttt{Period04} (see Fig. \ref{fig:f-4} b and Table \ref{fig:allpower}), and the eclipse depth curve is also fitted using Eq. \ref{eq:sine}, and the best fit is shown in Table \ref{tab:sine}.

The following results were obtained in this Section:
(a) The amplitude of NSH has a variation of 3.053(8) d.
(b) The variation of the eclipse depth is 3.040(6) d.
(c) O-C also has a periodic variation of 3.045(8) d.
The amplitude of the NSH, the eclipse depth, and the O-C all have a similar variation period to the SOR, which may provide new evidence for the precession of a tilted disk and will be discussed in Section \ref{sec:DISCUSSION}.

\begin{table*}
	\centering
	\caption{Results of the fit to O-C, eclipse depth and absolute value of the NSH amplitude using Eq. \ref{eq:sine}.\label{tab:sine}}
	\setlength{\tabcolsep}{1mm}{
		\begin{tabular}{llllllllllllllll} 
			\toprule
		Name&Start$^{\rm a}$ & End$^{\rm a}$ &  During& Z & Error & B & Error & A & Error &freq & Error &p & Error & $\nu$$^{\rm b}$ &$\chi^2$ $^{\rm c}$ \\ &[d] & [d] &  [d] & [mag] & [mag] &  [mag/d] &  [mag/d] & [mag] & [mag] &[1/d] & [1/d] & &  &[n]  & \\
			\midrule
		light&3418.5366&3469.4008&50.86&8.64&0.099&-0.003&0.00002865&0.1613&0.000598&0.32851&0.000039&0.83&0.14&58836&611.38\\
	$\mid A \mid$$^{\rm d}$&3418.8333&3469.2287&50.40&0.94&0.43&-0.000245&0.000125&-0.029&0.003&0.32774&0.00098&3.67&3.39&75&0.02\\
	Eclipse depth&3418.5530&3469.2900&50.74&2.50&1.18&-0.00051&0.000343&-0.10&0.01&0.32974&0.00075&-3.21&2.58&311&2.48\\
	O-C&3418.5530&3469.2900&50.74&0.0044&0.0037&-0.0000012&0.0000011&-0.00025&0.000022&0.32878&0.00098&-17.82&3.36&311&0.0000245\\
			\bottomrule
			
	\end{tabular}}
	\begin{threeparttable} 
		\begin{tablenotes} 
			\item [a] BJD2454833+. 
\item [b] Degree of freedom of the fit. 
\item [c] Fitted chi-square.
\item [d] The absolute value of the NSH amplitude.
		\end{tablenotes} 
	\end{threeparttable}
\end{table*}

\begin{table*}
	\centering
	\caption{Results of the segmented fit to the NSH using Eq. \ref{eq:sine}.\label{tab:k2(c18)}}
	\setlength{\tabcolsep}{2mm}{
		\begin{tabular}{lllllllllllllll} 
			\toprule
Start$^{\rm a}$ & End$^{\rm a}$ &  During& Z & Error & B & Error & A & Error &freq & Error &p & Error & $\nu$$^{\rm b}$ &$\chi^2$ $^{\rm c}$ \\ 
d & [d] &  [d] & [mag] & [mag] &  [mag/d] &  [mag/d] & [mag] & [mag] &[1/d] & [1/d] & &  &[n]  & \\
	\midrule
			3418.5366&3419.1128&0.5762&82.25&44.33&-0.024&0.013&0.067&0.060&6.876&0.044&-3480.99&151.67&653&1.74\\
			3419.1135&3419.7530&0.6395&-214.46&44.43&0.063&0.013&0.090&0.059&6.500&0.032&-2195.43&110.88&745&2.68\\
			3419.7537&3420.3932&0.6395&-55.21&41.34&0.016&0.012&0.135&0.053&6.372&0.020&-1758.47&66.79&746&2.74\\
			3420.3939&3421.0334&0.6395&172.40&40.52&-0.050&0.012&-0.118&0.053&6.475&0.022&-2112.07&75.03&714&1.61\\
			3421.0341&3421.6736&0.6395&-179.85&31.14&0.053&0.009&0.107&0.041&6.478&0.019&-2120.94&65.49&747&1.45\\
			3421.6743&3422.3138&0.6395&-311.77&32.56&0.091&0.010&0.070&0.043&6.856&0.030&-3416.31&102.31&747&1.36\\
			3422.3145&3422.9540&0.6395&-235.14&37.46&0.069&0.011&-0.117&0.050&6.560&0.021&-2401.68&72.84&746&2.09\\
			3422.9547&3423.5942&0.6395&173.87&38.09&-0.051&0.011&-0.125&0.051&6.552&0.020&-2376.97&69.78&746&2.09\\
			3423.5949&3424.2344&0.6395&562.19&32.36&-0.164&0.009&0.138&0.043&6.472&0.015&-2103.21&53.05&745&1.30\\
			3424.2351&3424.8746&0.6395&34.37&32.86&-0.010&0.010&0.119&0.044&6.465&0.018&-2079.99&63.00&725&1.60\\
			...&...&...&...&...&...&...&...&...&...&...&...&...&...&...\\
			\bottomrule
				
	\end{tabular}}
\begin{threeparttable} 
	\begin{tablenotes} 
		\item [a] BJD2454833+. 
		\item [b] Degree of freedom of the fit. 
		\item [c] Fitted chi-square.
		\item [Note] This table has 80 rows. Only ten rows are shown here. The rest of the data is in Table S1 of the supplementary material.
	\end{tablenotes} 
\end{threeparttable}
\end{table*}

\subsection{K2(c05)} 
\subsubsection{Positive Superhumps} \label{subsec:K2(c05)PSH}

No SOR was found in the light curve of K2(c05), so the locally weighted regression (LOESS; \citealp{cleveland1979robust}) fit was used instead of a sine fit.
After applying a LOESS fit with a span of 0.05 d to the out-of-eclipse curves (the solid blue circle of the curve \#1 in Fig. \ref{fig:k21} c), the residuals were taken for all curves (see the curve \#2 in Fig. \ref{fig:k21} c). In Section \ref{subsec:Frequencies}, a signal with a period of 0.174686(7) d is found in K2(c05), which can be identified as a PSH based on its relation to the orbital period, and the PSH excess ($\epsilon ^+ = (P_{\rm psh}-P_{\rm orb})/P_{\rm orb}$ ) was calculated to be $\epsilon ^- \sim$ 0.091. We first use the same segmental fitting method for the NSH to give a complete picture of the PSH. The data were divided into 117 segments of approximately 0.64 days each. However, the fit results were not ideal, with the best-fit frequencies becoming diffuse in the second half of the data (see Fig. \ref{fig:wwz} c). By checking on the light curve, the PSH may not be present in the second half of the data. To verify this, the Weighted Wavelet Z-transform (WWZ: \citealp{foster1996wavelets}) method was used, which can determine the frequency in the time domain.

We used WWZ on the detrended out-of-eclipse curve (the solid blue circle of the curve \#2 in Fig. \ref{fig:k21} c), and the results showed the presence of a clear signal around 0.174686(7) d, but the signal disappeared after approximately BJD2457178, a result that is almost identical to the segmented sine fit (see Fig. \ref{fig:wwz} c and d). To verify this further, we performed separate frequency-amplitude analyses of the out-of-eclipse curves before and after BJD2457178, which showed the presence of a significant PSH before BJD245717 but its complete disappearance after BJD245717 (see Fig. \ref{fig:wwz} a and b). This study demonstrates that PSH is not a permanent presence in SDSS J0812.

\subsubsection{Eclipse depth and O-C} \label{subsec:PSH}

\begin{figure*}
	\includegraphics[width=1.5\columnwidth]{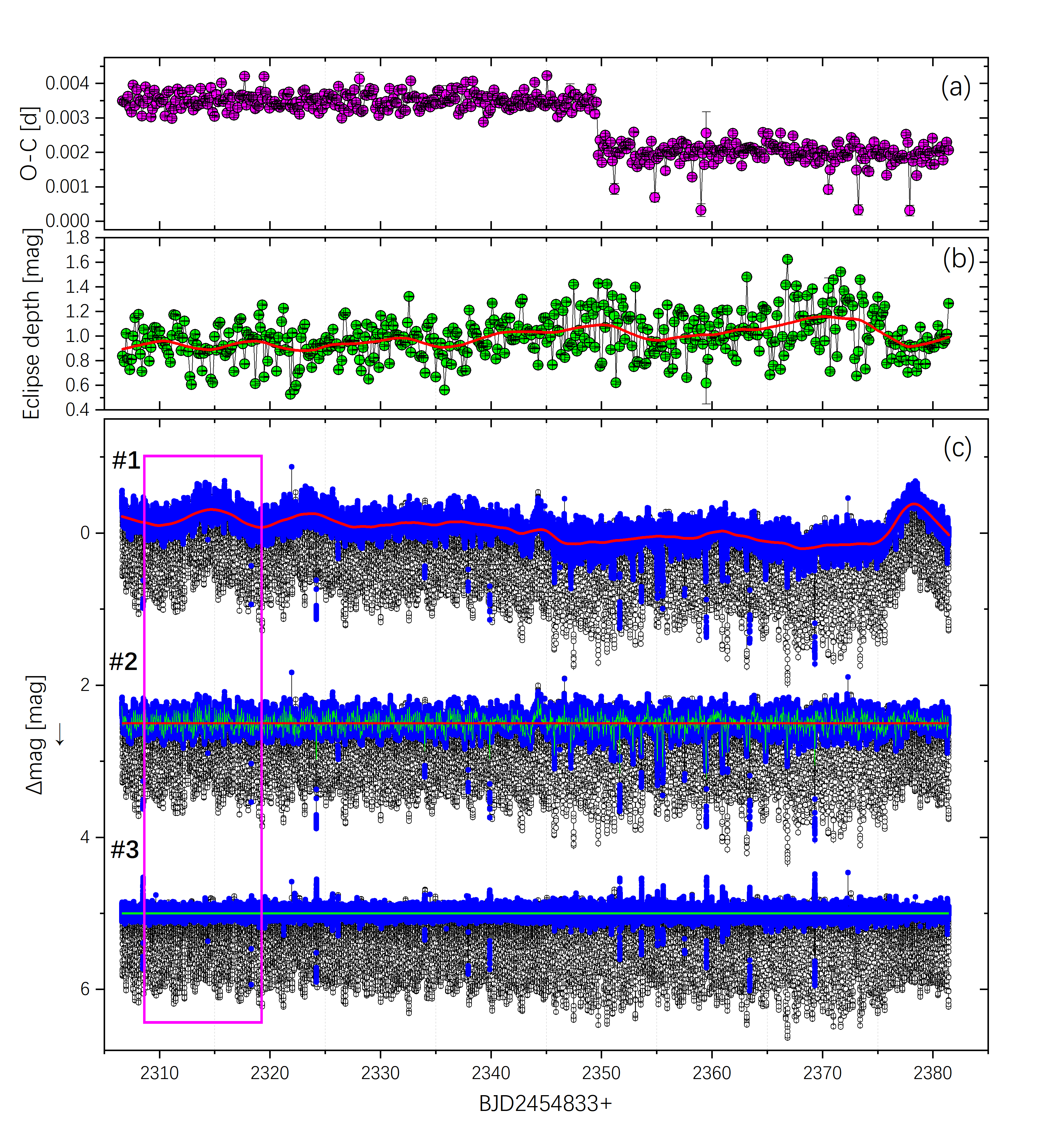}
	\caption{A combined plot of the analytical process and results for K2(c05). (a): O-C plots; (b): plots of the eclipse depth; (c): we split into 3 main curves; \#1: all light curves of K2(c05), the solid blue circle is the out-of-eclipse part, the red curve is the LOESS fit with a span of 0.05 d to the out-of-eclipse curve; \#2: the residuals of the LOESS fit to all curves plus 2.5 mag, the green curve is the LOESS fit with a span of 0.001 d to the out-of-eclipse curve; \#3: the residuals of the second LOESS fit to all curves plus 5 mag.\label{fig:k21}}
\end{figure*}

\begin{figure*}
	\includegraphics[width=1.5\columnwidth]{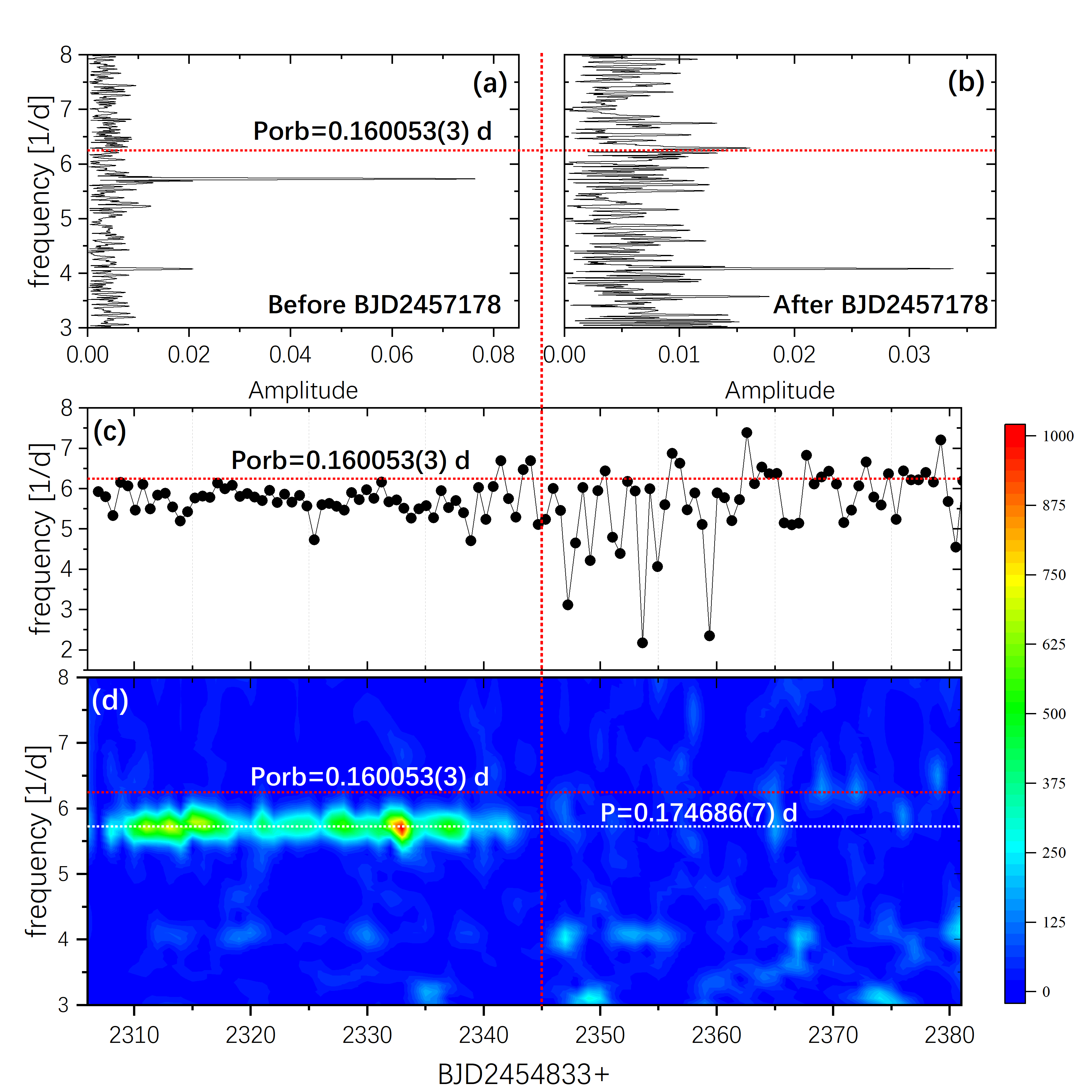}
	\caption{Frequency-Amplitude plot of the out-of-eclipse curves in K2(c05), segmented linear superimposed sinusoidal fit and WWZ 2d spectrum. (a) and (b) are the frequency-amplitude spectrum for the out-of-eclipse portion of the \#2 curve less than and greater than BJD2457178 in Fig. \ref{fig:k21} c, respectively. (c): frequencies obtained by linearly superimposing a sinusoidal fit to the segmented out-of-eclipse curve of the \#2 curve in Fig. \ref{fig:k21} c. (d): The WWZ 2d spectrum of the out-of-eclipse curve of \#2 curve in Fig. \ref{fig:k21} c. The horizontal red and white dotted lines in the plot are 0.160053(3) d and 0.174686(7) d, respectively, and the vertical dotted line in the plot is BJD2457178. \label{fig:wwz}}
\end{figure*}

\begin{figure}
	\includegraphics[width=\columnwidth]{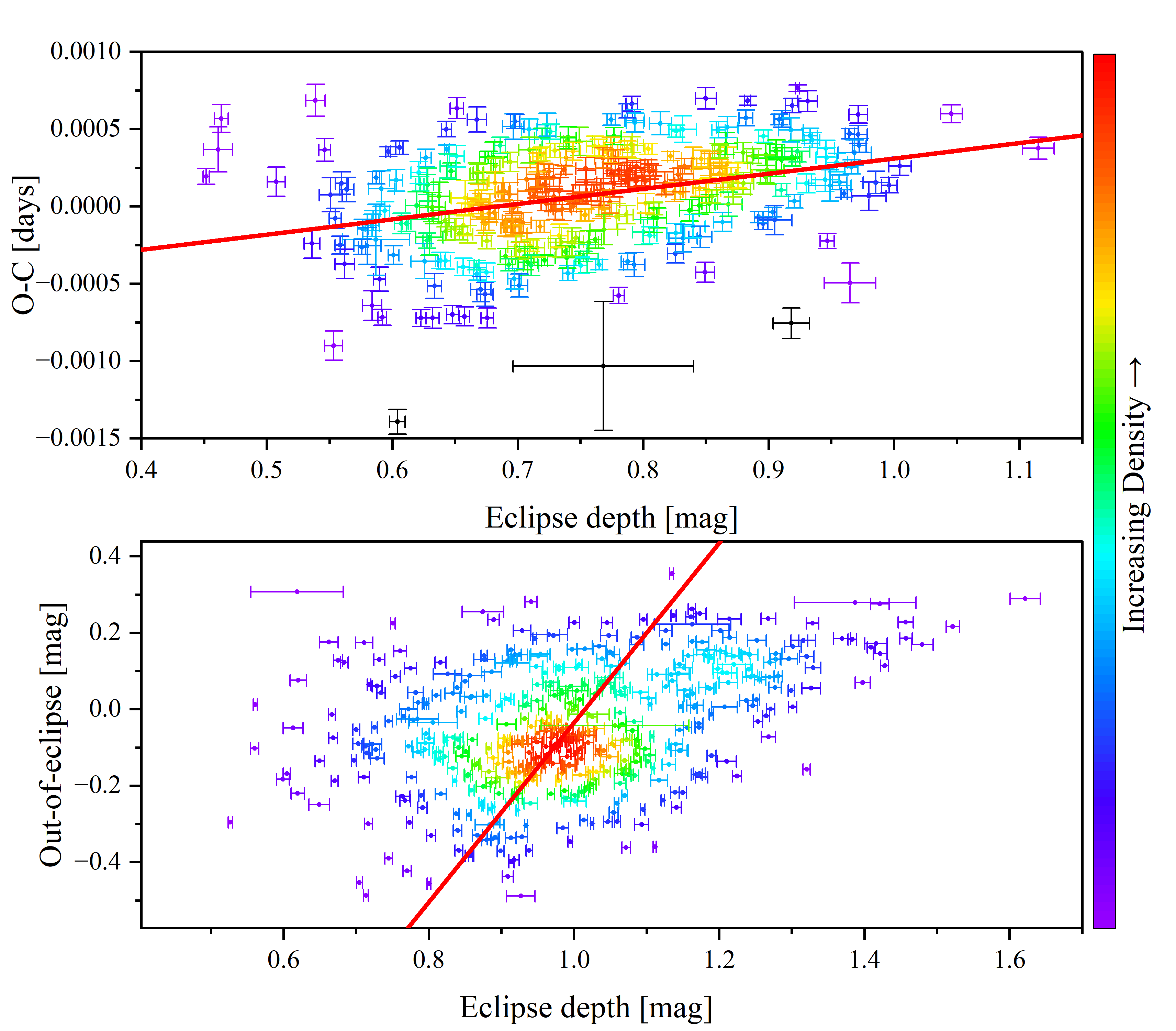}
	\caption{The relationship between the eclipse depth and the O-C and mean magnitude out-of-eclipse, respectively. (a) Relationship between eclipse depth and O-C in K2(c18); (b) Relationship between eclipse depth and mean magnitude out-of-eclipse in the K2(c05) light curve. The solid red line in the figure is a linear fit without weight. \label{fig:dep-out-o-c}}
\end{figure}

The PSH caused the out-of-eclipse curve to be variable, but PSH was not persistent, so our method of using a segmented sine fit was unsuccessful (see Fig. \ref{fig:wwz} c). We used a LOESS fit with a span of 0.001~d instead of the segmented sinusoidal fit to remove the PSH trend (see a solid green line of the curve \#2 in Fig. \ref{fig:k21} c). Eclipse depths and minima were calculated using the same method as in Section \ref{subsec:O-Ck22} for the light curve with the PSH trend removed (the curve \#3 in Fig. \ref{fig:k21} c). Smoothing of the eclipse depth reveals a possible opposite trend to the original light curve. To further verify this, the average magnitude of the original out-of-eclipse curve was calculated and plotted versus eclipse depth and fitted linearly (see Fig. \ref{fig:dep-out-o-c} b). The results further confirm that the brightness out-of-eclipse decreases as the eclipse depth becomes deeper.

\begin{figure}
	\includegraphics[width=\columnwidth]{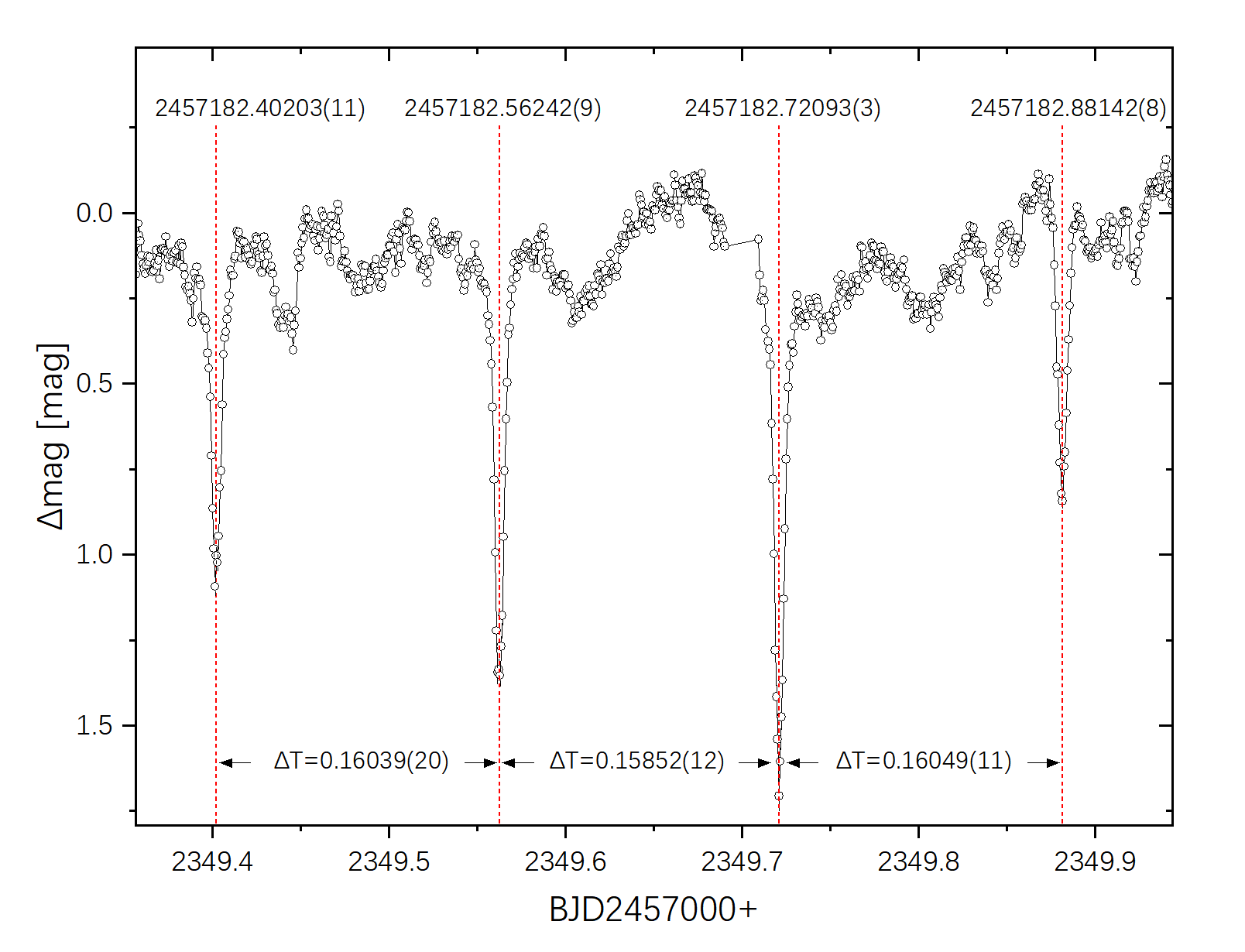}
	\caption{Four eclipse curves for curve \#1 in Fig. \ref{fig:k21} (c). The vertical red dotted lines are the minima corresponding to each eclipse; the minima are marked at the top of the dotted lines, and the time difference between each minima is indicated by $\Delta$T between the vertical lines.\label{fig:tiao}}
\end{figure}

A total of 465 minima were obtained at K2(c05) by Gaussian fitting, and O-C analyses were performed using ephemeris \ref{eq:ephemeris}. The O-C results show that the minima in K2(c05) do not suffer from oscillations similar to those in K2(c18), but there is a jump between the minima BJD2457182.56242(9) and BJD2457182.72093(3) (see Fig. \ref{fig:k21} a). A change of $\sim$ 0.0015 d ($\sim$ 130 s) in one cycle is impossible. To confirm this change, we found the corresponding part of the eclipse on the light curve (see Fig. \ref{fig:tiao}). The analysis showed that the jump variation was only present between BJD 2457182.56242(9) and 2457182.72093(3) and that there were no anomalies in the eclipse before or after them. This is further supported by the fact that there is no significant change in the slope of the O-C plot before and after the jump. There is no strong reason to prove that the stars are different before and after the jump, because the period before and after the jump did not change; the PSH disappearance occurred at about BJD2457178, which had disappeared before the jump (BJD2457182.56242(9)) and was not related to the jump. Although the jump is close to twice the exposure time, there is no precedent nor can we conclude that it comes from systematic errors. Such a jump is very odd and we suggest that both physical causes and observational system errors deserve to be concerned.

\begin{figure}
	\includegraphics[width=\columnwidth]{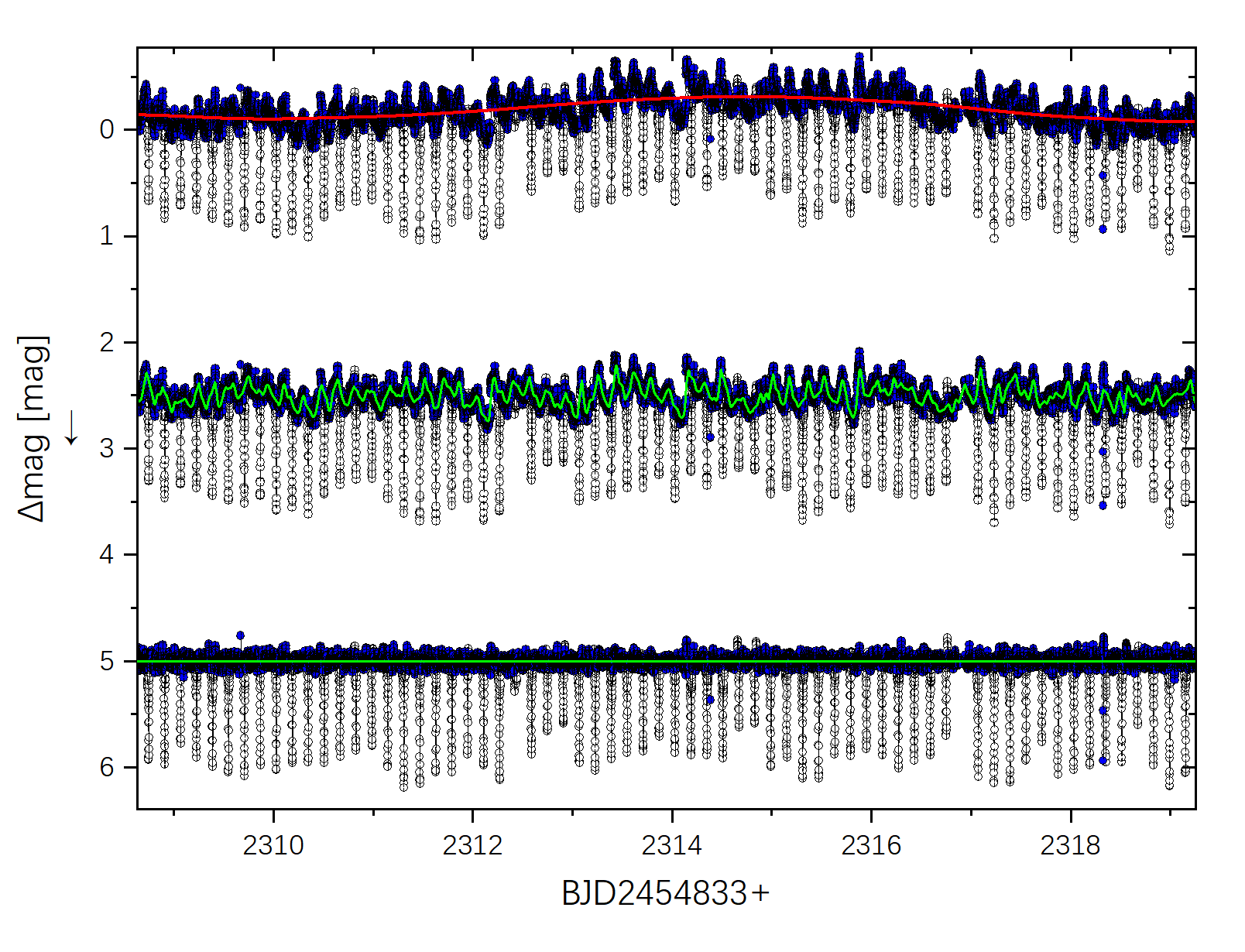}
	\caption{Enlarged view of the magenta rectangular box of the (c) plate in Fig. \ref{fig:k21}.\label{fig:k21-2}}
\end{figure}

\subsection{\textit{TESS}} 

\begin{figure}
	\includegraphics[width=\columnwidth]{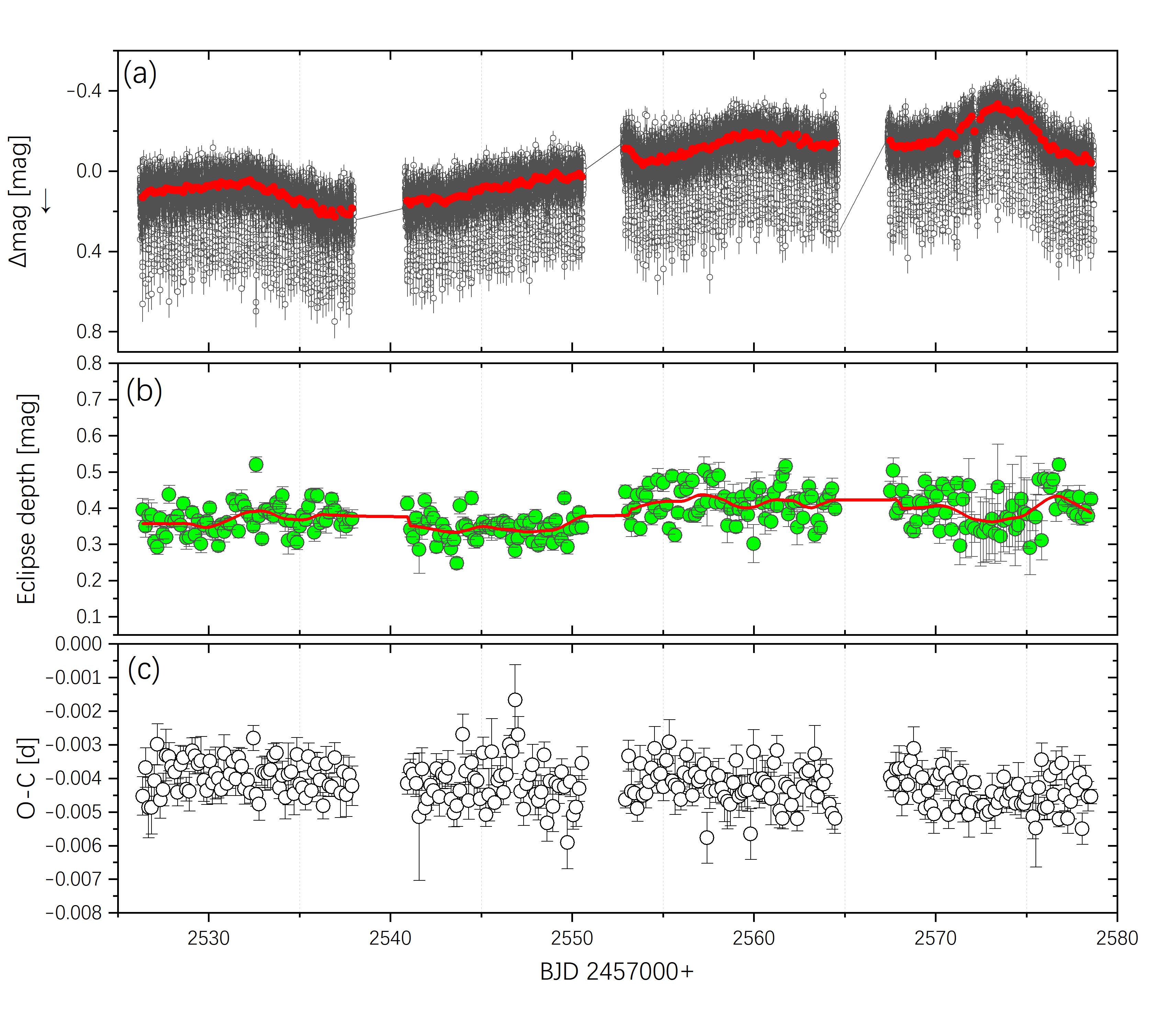}
	\caption{light curve and results for \textit{TESS}. (a): Light curve of \textit{TESS}, the solid red circles are the out-of-eclipse mean curve corresponding to each eclipse; (b): eclipse depth variation curve, the solid red line is the LOESS fit; (c): O-C plot. \label{fig:tess}}
\end{figure}

\textit{TESS} likewise observed SDSS J0812, and since no anomalies were found other than the orbital period signal, we left the trend alone and used the same method as in Section \ref{subsec:O-Ck22} to calculate the eclipse depth and minima. A total of 275 minima were obtained. The results show that the correlation between the eclipse depth and the eclipse profile is insignificant, with an average eclipse depth of $\sim$ 0.38 mag. The O-C analysis shows no significant oscillations or jumps (see Fig. \ref{fig:tess}). The absence of SOR, NSH and PSH in the \textit{TESS} photometry proves that none are permanently present in the SDSS J0812.

\subsection{Peculiar signal} \label{subsec:outbursts}

\begin{figure}
	\includegraphics[width=\columnwidth]{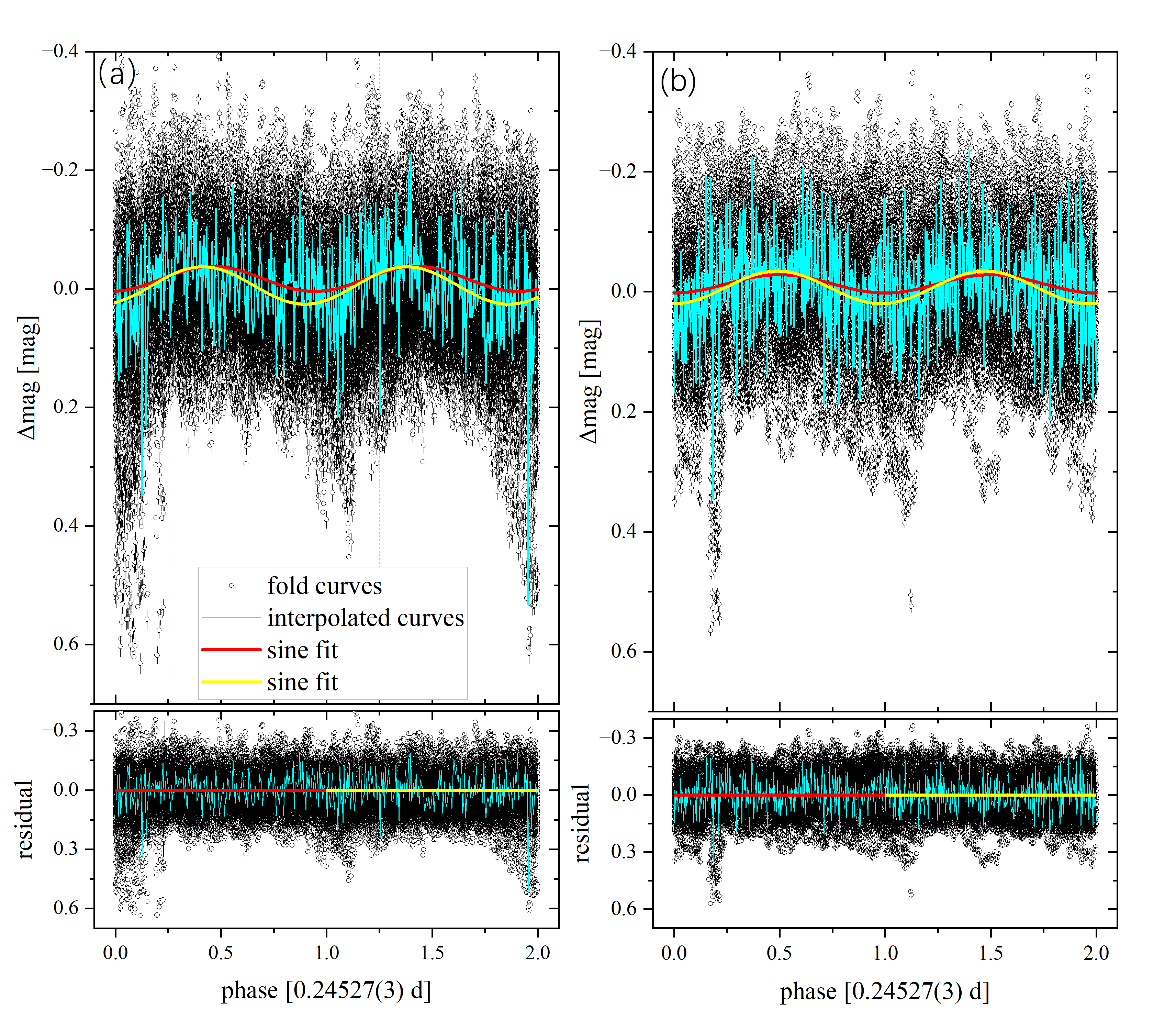}
	\caption{Folding curves for K2(c05) and K2(c18) with 0.24527 d. (a) and (b) are folding curves using 0.24527 d as the period for the out-of-eclipse curve in K2(c05) and K2(c18), respectively. The solid red line is a sine fit to the original folding curve, and the solid yellow line is a sine fit after linear interpolating the folding curve.\label{fig:0.24527d}}
\end{figure}

There is a peculiar signal of 5.89 hr in both K2(c05) (P = 0.24526(2) d, A = 0.027(1) mag) and K2(c18) (P = 0.24527(3) d, A = 0.016(1) mag) (see Fig. \ref{fig:allpower} and Fig. \ref{fig:wwz} a and b). It is difficult to find on the light curve due to the small amplitude. Therefore we use a folding approach for verification. Because the signal difference between the two Campaigns was minimal, 0.24527(3) d was used uniformly for folding.
The data in part K2(c05) were folded using a 0.24527(3) d using the out-of-eclipse curve (the solid blue circle of the curve \#2 in Fig. \ref{fig:k21} c).
The results of the sinusoidal fit of the folded curve indicate that this periodic signal is present (see Fig. \ref{fig:0.24527d} a). We reduced the number by performing a linear interpolation of 500 points on the folded curves. The periodic signal became more apparent when a sinusoidal fit was performed on the interpolated curve. We similarly folded the out-of-eclipse curve with the de-SOR and NSH (sinusoidal part only) trends in K2(c18), again using the same number of interpolation points (see Fig. \ref{fig:k21} b). A sinusoidal fit to the interpolated curves shows that the signal is also present. The folded curve and Fourier plot verified that the 5.89 hr signal might be present in SDSS J0812 but was not successfully found using a sinusoidal fit to the time series data, probably due to the small amplitude. Therefore, the 5.89 hr periodic signal still requires further observation and verification.

\section{DISCUSSION} \label{sec:DISCUSSION}

\subsection{Parameter calculation} \label{subsec:Parameter calculation}

We estimated the mass of the secondary star to be $M_2=0.350M_\odot$ based on the orbital period 0.160053(3) d and semi-empirical mass-period relationship ($M_{2} = 0.065P _{\rm orb} (hr)^{1.25}$; \citealp{warner2003cataclysmic}).
The mass ratio is estimated to be in the range to be $0.243 < q < 0.833$, based on the mass ratio $q = \frac{M_{2}}{M_{1}} \le \frac{5}{6}$ required for stable mass transfer in CVs, the mass of the white dwarf is less than the Chandrasekhar limit ($M_1 < 1.44 M_\odot$). 

In addition, the range of mass ratios can be further narrowed down based on $\epsilon ^+ \sim 0.091$ and  $\epsilon ^- \sim$ 0.050. Based on \cite{2006MNRAS.371..235P} gives the empirical formula for fitting the mass ratio to the $\epsilon ^+$:
\begin{equation}
	\epsilon^{+} =- 4.1\times10 ^{-4} + 0.2076q
\end{equation}
obtaining $q$ = 0.44. The same can be calculated using the relation of the $\epsilon ^-$ to $q$ given by \cite{wood2009sph}:

\begin{equation}
	q = -0.192\mid\epsilon^{-}\mid ^{1/2} + 10.37\mid\epsilon^{-}\mid - 99.83\mid\epsilon^{-}\mid ^{3/2} + 451.1\mid\epsilon^{-}\mid ^{2}
\end{equation}
$q$ is calculated as 0.488.
Combining these two calculations, a mass ratio of 0.465(23) is obtained, so that $M_1=0.750(37)M_\odot$. The separation of the binary is obtained based on Kepler's law as $a=1.284(14)R_\odot$.
According to \cite{Paczynski1977} and \cite{Lasota2008} the maximum accretion disk radius can be expressed as:
\begin{equation}
	\frac{R_{\rm d}(\rm max)}{a}=  \dfrac{0.60}{1+q}		
\end{equation}
Therefore the maximum accretion disk radius of SDSS J0812 is obtained as $R_{\rm d}(\rm max)=0.526(32) R_\odot$.

\begin{figure}
	\includegraphics[width=\columnwidth]{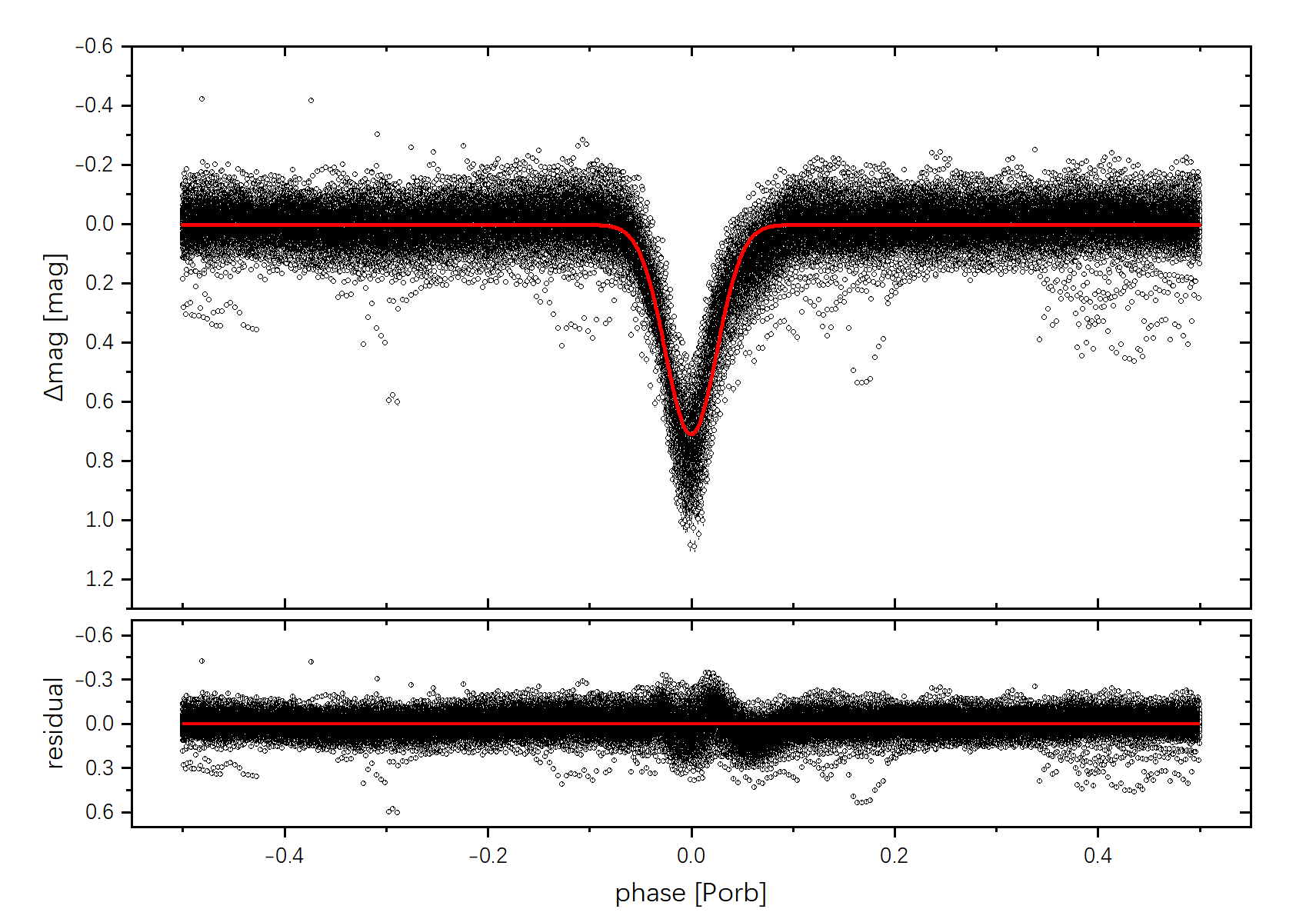}
	\caption{Folded diagram of the eclipse. Data are light curves of K2(c18) with SOR and NSH removed, with a folding period of 0.160053(3) d. The solid red line is the Gaussian fit.\label{fig:Folded}}
\end{figure}

The relationship between $q$ and $i$ was given by \cite{dhillon1991spectrophotometric} and \cite{eggleton1983approximations}:

\begin{equation}
	\left(\frac{R_{2}}{a}\right)^{2} = \sin^{2}(\pi \Delta \phi_{1/2}) + \cos^{2}(\pi \Delta \phi_{1/2})\cos^{2}i 
\end{equation}
and
\begin{equation}
	\left(\frac{R_{2}}{a}\right) =  \dfrac{0.49q^{2/3}}{0.6q^{2/3}+\ln(1+q^{1/3})}		
\end{equation}
$a$ is the binary separation, and $\Delta\phi_{1/2}$ is the phase full-width of the eclipse at half the out-of-eclipse.
We used K2(c18) to detrend the data for SOR and NSH, folded by 0.160053(3) d (see Fig. \ref{fig:Folded}). A Gaussian was fitted to the folding curves, and the full width at half maximum (FWHM) was 0.0596(1). we used the FWHM as the phase full-width of the eclipse at half the out-of-eclipse ($\Delta\phi_{1/2}$ = 0.0596(1)). Therefore, combined with the  $q$, the orbital inclination is limited to $i = 75.03 (\pm0.30)$.

\subsection{Three windows for studying tilted disk} \label{subsec:Three windows}

In Section \ref{subsec:K2(c18)1}, periodic variations in the O-C (3.045(8) d), eclipse depth (3.040(6) d), and amplitudes (3.053(8) d) of NSH were found to be broadly consistent with the SOR (3.0451(5) d), and the difference between them is only about 0.01 d. All three of these periodic signals are likely associated with the tilted disk, providing us with a new window to study the tilted disk. 
The periodic variation of O-C represents the periodic variation of the position of the brightness center in the primary system. Hot spots and accretion streams are significant players in the formation and evolution of accretion disks, and current studies suggest that they are also essential factors in the formation and evolution of tilted disks. Therefore, variations in brightness centers are highly likely associated with tilted disks, hot spots, and accretion streams.

The eclipse depth was also found to have a variation cycle consistent with the SOR. The eclipse depth characterizes the brightness of the eclipse center. The brighter the center, the deeper the eclipse. The relationship between eclipse depth and O-C was plotted and linearly fitted, and the results demonstrated that O-C increased with increasing eclipse depth (see Fig. \ref{fig:dep-out-o-c} a). The combination of the O-C variation suggests that the brightness center of the accretion disk system not only varies periodically in position but also in brightness and that the variation period is consistent with the tilted disk precession cycle. 
The amplitude of the NSH was likewise found to have a period of variation essentially consistent with the SOR, which directly confirms that the origin of the NSH is closely related to the tilted disk precession. 
Thus the variations of O-C, eclipse depth, and NSH amplitude provide three valuable windows for studying the tilted accretion disk.

\begin{figure*}
	\includegraphics[width=1.5\columnwidth]{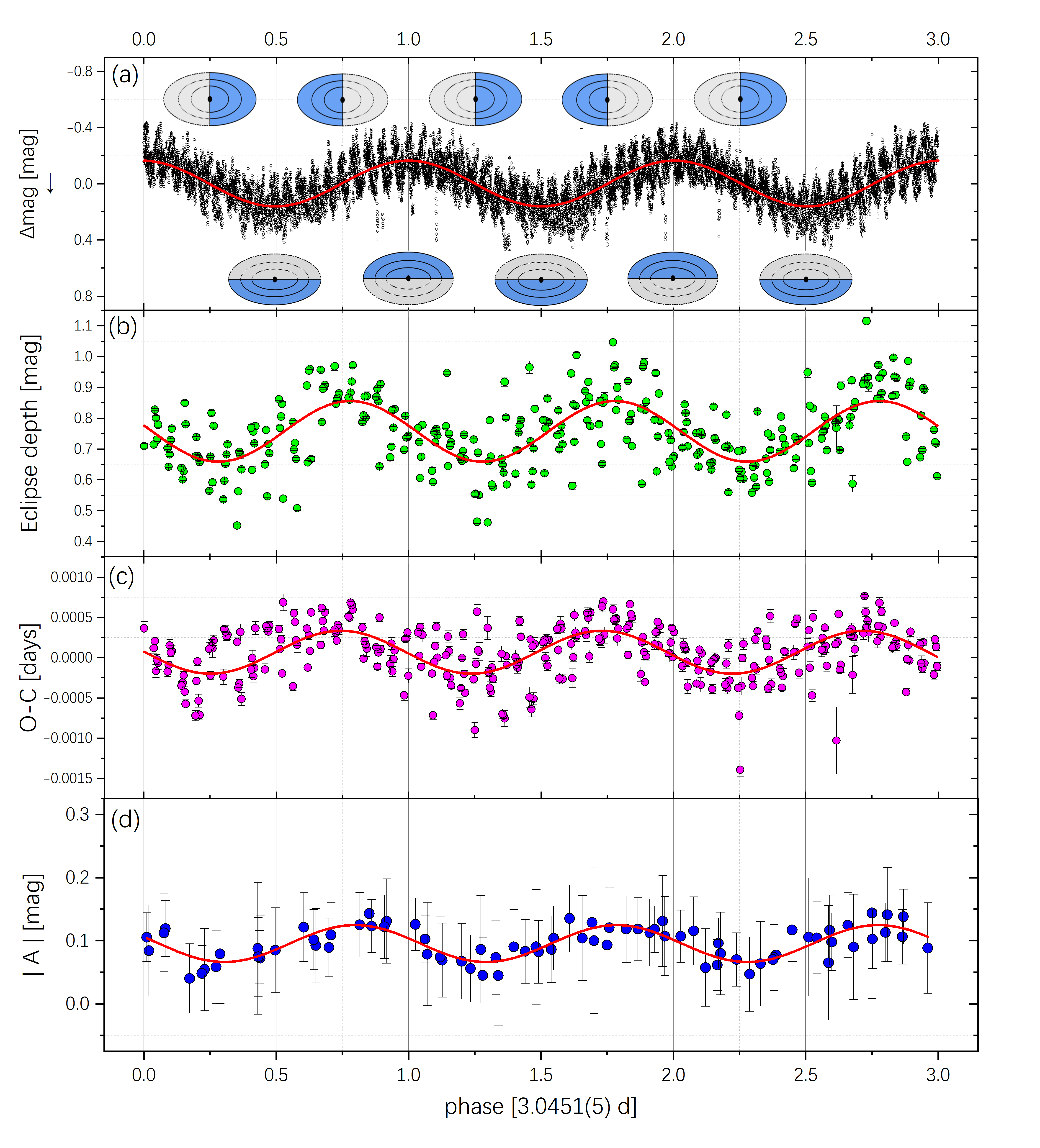}	\caption{Folding curves using 3.0451(5) d for different curves in the K2(c05). (a): folded curves of the light curve corresponding to the out-of-eclipse part of curve \#1 in Fig. \ref{fig:all_light_k22} d; the ellipse in the panel (a) is a hypothetical retrograde precession of the tilted accretion disk without parameters, the black dot in the center represents the white dwarf, the solid blue area represents the semi-tilted disk above the orbital plane, the gray dashed area represents the semi-tilted disk below the orbit, the solid line where the gray and blue intersect is the nodal line, the disk precession clockwise, the solid or dashed line pressed in the center of the disk is its corresponding precession phase (SOR phase); (b), (c) and (d) are folded curves of the eclipse depth, O-C and the absolute values of NSH amplitudes, respectively, corresponding to (b), (a) and (c) of Fig. \ref{fig:all_light_k22}. \label{fig:so-oc-dep-k-2-5}}
\end{figure*}

We have assumed a reverse (clockwise) precession of the tilted disk without considering the parameters at 0, 0.25, 0.75 and 1 precession phases ($\phi_{\rm prec}$) respectively (see Fig. \ref{fig:so-oc-dep-k-2-5} a). The view of the observer is perpendicular to the figure. The following discussions all default to a counterclockwise rotation of the binary star around, with the hot spots in the right semi-disk when $\phi_{\rm orb}$ = 0.
Taking the orbital plane as the reference when divided by the orbital plane can divide the semi-disk below or above the orbital plane into the upper and lower halves. Dividing the tilted disk from the observer's perspective, the accretion disk can be divided into the left and right semi-disk.

To show more intuitively the relationship between O-C, eclipse depth, NSH amplitude, and SOR, we folded them simultaneously with the maximum value of SOR (BJD2458275.58251) as $\phi_{\rm prec}$ = 0 with a period of 3.0451(5) d (see Fig. \ref{fig:so-oc-dep-k-2-5} ).
By comparing the folding results, it is surprising to find that there is a consistent phase change in the O-C, eclipse depth, and NSH amplitudes, while the maxima and minima are reached at $\phi_{\rm prec}$ $\sim$ 0.75 and  $\phi_{\rm prec}$ $\sim$ 0.25, respectively. This important finding may provide new observational evidence for the precession of the tilted accretion disk.

The SOR is considered a periodic light change caused by the tilted accretion disk precession, which is gradually recognized.
The NSH is thought to have originated from the interaction of the tilted disk reverse precession with orbital motion, but the exact origin is controversial.
In our work, we found for the first time that the amplitude of the NSH exists at a consistent variation period of the SOR but also found that the amplitude of the NSH occurs maxima at $\phi_{\rm prec}$ $\sim$ 0.75. This provides evidence for a direct relationship between the origin of the NSH and the precession of the tilted disk.

\subsection{Change of eclipse center} \label{subsec:eclipse center}	

The magnetic activity cycle in the secondary star can be used to explain the periodic change revealed by O-C, but the time scale of this mechanism is about ten years or more \citep{Applegate1992mechanism}. The light travel-time effect caused by the existence of the third body is also used to explain the periodic change of the orbital period revealed by O-C, but it also belongs to the long-term scale change (\citealp{irwin1952determination}; \citealp{borkovits1996invisible}).
The material transfer of binary stars discussed above is also a long-term change. 
At present, some studies believe that the slight amplitude jitter of the accretion disk will cause mid-eclipse times to advance or delay in the observation \citep{schaefer2021fast}. When a jitter occurs before the eclipse, it will lead to the advance of mid-eclipse times. When a jitter occurs after the eclipse, it will cause a delay in mid-eclipse times. However, the jitter is random, and the timescale is in years\citep{schaefer2021fast}. 
\citet{vogt1982z} proposed an eccentric accretion disk model to explain SU UMa's superoutburst phenomenon. The existence of an eccentric accretion disk can change the brightness center of the accretion disk, resulting in a change in the observed mid-eclipse times. Therefore, the period of O-C periodic oscillation is between the orbital period and the superhump period (\citealp{vogt1981eclipsing}, \citealp{krzeminski1985eclipsing}). The O-C oscillation caused by the eccentric accretion disk model is strictly periodic, and the period is short.
This phenomenon that the mid-eclipse times at the time of outburst is earlier than quiescence was found in V729 Sgr \citep{ramsay2017v729}, V447 Lyr \citep{ramsay2012kepler} and KIS J192748.53+444724.5 \citep{scaringi2013kepler}. They reasonably explained that the brightness fraction of the hot spots in quiescence is significant, and the brightness fraction of the hot spots decreases when it bursts, so the mid-eclipse times are earlier than that in quiescence.

We found periodic variations in the O-C of SDSS J0812 with periods and amplitudes of 3.045(8) d and 0.00026(2) d ($\sim$ 22 s), respectively.
\citet{Miguel2016accretion} likewise found that the O-C of the nova-like Ursa Major of the NSH system has a SOR-consistent periodic variation of $3.68$ d with an amplitude of $\pm 33$ s, which they also attribute to the periodic variation of light center of the accretion disk.
This demonstrates that the O-C in the NSH system with changes consistent with the SOR cycle is no exception. A series of studies have shown that the eclipse centers of some CVs are unstable and that accretion disk and hot spots may be responsible for the variation in eclipse centers.

In the theory of \cite{Barrett1988MNRAS}, \cite{Patterson1997PASP}, \cite{Wood2007ApJ...661.1042W}, and \cite{Montgomery2012}, the hot spots and the accretion stream are the primary sources of variation under the assumption of the tilted disk precession. Whether the stream enters the disk directly or the particles splashed from the edge of the disk by the stream impact enter the inner ring, in any case, the change in the accretion stream or the change in the position of the hot spots causes the upper or lower disk to brighten periodically. However, since the disk is optically thick, the observer can only see the brightness modulation once in an orbital phase. \cite{Montgomery2012}'s simulation suggests that the particles can only splash into the disk on the side close to the secondary star, and therefore which side is brightened and which is observed is related to the phase of the tilted disk and the orbital phase of the binary.

Our entry point is the eclipse, so we discuss the new phenomenon by the relationship between the phase of the tilted disk, the hot spots, and the position of the accretion stream when the eclipse ($\phi_{\rm orb}$ = 0). 
When $\phi_{\rm orb}$ = 0 and $\phi_{\rm prec}$ = 0.25, the right semi-disk is above the orbital plane, and the accretion stream touches the right semi-disk, the back side of the right semi-disk will become brighter but cannot be observed, when $\phi_{\rm orb}$ = 0 and $\phi_{\rm prec}$ = 0.75, the right semi-disk is below the orbital plane, and the accretion stream touches the right semi-disk, the front side of the right semi-disk will become brighter and can be observed. Thus, such a variation explains the shift of the eclipse center: it appears to the observer that the brightness distribution is relatively uniform over the disk when $\phi_{\rm orb}$ = 0 and $\phi_{\rm prec}$ = 0 or 0.5. However, when $\phi_{\rm orb}$ = 0, $\phi_{\rm prec}$ = 0.75, and $\phi_{\rm prec}$ = 0.25, there is a significant shift in the center of brightness, thus causing the minima of the eclipse to be advanced or delayed with the tilted disk precession.

To further verify the movement of the center of the eclipse, we estimate the extent of the movement. We idealize the motion of the binary stars by assuming that the secondary star moves in a circular motion around the primary star. Using the primary star as the reference system, the angular velocity of the secondary new can be estimated based on $P_{\rm orb} = 0.160053(3)$ d and $a = 1.284(14) R_\odot$ as $\omega$ = 2249.25(4) $^{\circ}$/d. Combining the amplitude (A = 0.00026(2) d) of the O-C and the $\omega$ and the trigonometric formula, we obtain the range of variation of the eclipse center as 0.709(57) $R_\odot$. The range of variation is smaller than the diameter ($1.052(64) R_\odot$) of the maximum accretion disk, so the scale of the brightness center shift is possible.

The minimum and maximum eclipse depths are also at $\phi_{\rm prec}$ = 0.25 and $\phi_{\rm prec}$ = 0.75, respectively, rather than at the overall brightness maximum and minimum of $\phi_{\rm prec}$ = 0 and  $\phi_{\rm prec}$ = 0.5, which is confusing. We have found that the O-C and the eclipse depth vary similarly, with the eclipse depth increasing when O-C increases (see Fig. \ref{fig:dep-out-o-c} a and Fig. \ref{fig:so-oc-dep-k-2-5}), suggesting that the change in the eclipse depth may be related to the shift of the brightness center.

Based on the analysis of the eclipse position and the eclipse depth, our explanation may prove that the accretion disk is reverse precession. This is because only when $\phi_{\rm orb}$ = 0 and $\phi_{\rm prec}$ = 0.75 or $\phi_{\rm prec}$ = 0.25, the lower semi-disk is on the right side is on the same side as the accretion stream and hot spots. If the accretion disk is forward-precession, then the maxima occurrence of O-C and the eclipse depth is in $\phi_{\rm prec}$ = 0.25. Our conclusion may prove that the accretion disk is reverse precession.

\section{CONCLUSIONS} \label{sec:CONCLUSIONS}

Based on the photometry of \textit{K2} and \textit{TESS}, we investigate the super-orbital signals, negative superhumps, positive superhumps, O-C, eclipse depth, and negative superhumps amplitude of the eclipsing cataclysmic variable star SDSS J0812. The summary of this paper is as follows.

(1) For the first time, we found the super-orbital signals with 3.0451(5) d in SDSS J0812. We also found negative superhumps with period and amplitude of 0.152047(2) d and 0.0891(8) mag, respectively, and the excess was calculated to be $\epsilon ^- \sim$ 0.050. A least-squares fit by dividing the out-of-eclipse curve into 80 segments with a linear superposition of sines is performed separately. Based on the relevant results, we found for the first time that the negative superhumps amplitude has a periodic variation with periods and amplitudes of 3.053(8) d and 0.029(3) mag, respectively, and found that the amplitude reaches a maximum at $\phi_{\rm prec} \sim $ 0.75, which provides new evidence for the precession of the tilted disk.

(2) Based on the K2(c18) light curve with the super-orbital signals and the superhumps trend removed, 316 minima were obtained using a Gaussian fit. By analyzing the O-C of the minima, we found that there are periodic variations of the O-C with periods and amplitudes of 3.045(8) d and 0.00026(2) d ($\sim$ 22 s), respectively, again reaching a maximum at $\phi_{\rm prec}$ $\sim$ 0.75. We suggest that the variation of the brightness center of the incoming tilted disk causes O-C variation. We estimate the range of variation of the brightness center to be 0.709(57) $R_\odot$, which is smaller than the maximum disk diameter, suggesting that the variation of the brightness center is possible. In addition, we calculated the eclipse depth of SDSS J0812 and found the same periodic variation with periods and amplitudes of 3.040(6) d and 0.096(6) mag, respectively, which also reaches a maximum at $\phi_{\rm prec}$ $\sim$ 0.75. Based on the variation of the eclipse depth and O-C, it shows that not only is there a periodic variation in the position of the center of the disk brightness of SDSS J0812, but also the brightness varies with the position.
The periodic variation of the eclipse depth likewise supports the precession of the tilted disk.

(3) Similarly, we performed a similar analysis on the light curve of K2(c05). Based on the sine fit,  \texttt{Period04}, and WWZ methods, we found for the first time that SDSS J0812 has positive superhumps with periods and amplitudes of 0.174686(7) d and 0.035(1) mag, respectively. However, the positive superhumps exist only before BJD2457178, which proves that the positive superhumps are not permanently present in SDSS J0812. We obtained 465 minima based on detrended light curves and Gaussian fits. The results of the O-C analysis suggest that the oscillation case in K2(c18) does not exist. The same computed eclipse depth also did not find oscillations, but an opposite evolutionary trend of the eclipse depth and the out-of-eclipse brightness of the original curve was found, providing a new phenomenon for the study of the nova-like object.

(4) By analyzing the \textit{TESS} photometry, we obtained 275 minima. O-C and eclipse depth curves indicate the disappearance of the oscillations like the super-orbital signals. Also, the super-orbital signals, negative superhumps, and positive superhumps all disappear in the \textit{TESS} photometry, which confirms that none of them is permanently present in SDSS J0812. The theory of the line of apsides precession and nodal precession of the accretion disk faces not only the test of how it was generated but also the question of why it disappeared. 

(5) Finally, we found the signal with 5.89 hr in K2(c05) and K2(c18) and verified by folding curves, but challenging to find at specific times, so further observations and verification are needed. Positive superhumps and negative superhumps exist in K2(c05) and K2(c18), respectively, but at the same time, there is a 5.89 hr signal, and we suggest that the signal in SDSS J0812 may be a way to study the association between the negative superhumps and positive superhumps.

\section*{Acknowledgements}

This work was supported by the National Natural Science Foundation of China (Nos. 11933008 and 12103084). Data from \textit{TESS} and \textit{K2} were used in this work, detailed in Section \ref{sec:style}, and we thank all the staff for \textit{K2} and \textit{TESS} for making our study possible.

\section*{Data Availability}
The corresponding author will share all other data underlying this article on reasonable request.



\bibliographystyle{mnras}
\bibliography{example} 

\bsp	
\label{lastpage}
\end{document}